\documentclass[11pt,letterpaper]{article}

\usepackage[margin=1in]{geometry}
\usepackage{microtype}

\usepackage{amsmath,amsfonts,amssymb,amsthm}
\usepackage{thmtools}

\usepackage{tikz}
\usetikzlibrary{arrows.meta}
\usepackage{subcaption}
\usepackage{multirow}

\usepackage{xcolor}
\usepackage[
    hypertexnames=false,
    colorlinks=true,
    linkcolor=blue,
    citecolor=blue,
    urlcolor=blue
]{hyperref}

\newtheorem{theorem}{Theorem}[section]
\newtheorem{lemma}[theorem]{Lemma}

\newtheorem{definition}[theorem]{Definition}
\newtheorem{example}[theorem]{Example}
\newtheorem{remark}[theorem]{Remark}

\newenvironment{proofof}[1]{\begin{proof}[Proof of #1]}{\end{proof}}

\newcommand{\mini}[1]{\mbox{minimize} & {#1} &}
\newcommand{\maxi}[1]{\mbox{maximize} & {#1} &}
\newcommand{\st}{\mbox{subject to} }
\newcommand{\con}[1]{&#1 &}
\newcommand{\qcon}[2]{&#1, & \mbox{for } #2.}
\newenvironment{lp}{\begin{equation}\begin{array}{lll}}{\end{array}\end{equation}}

\newcommand{\eps}{\epsilon}
\newcommand{\best}{\text{Best-}\eps\text{-NE}}
\newcommand{\worst}{\text{Worst-}\eps\text{-NE}}

\tikzset{
    node/.style={circle, draw=black, minimum size=8mm, inner sep=0pt},
    edge/.style={draw=black}
}

\title{How Much Can Approximate Equilibria\\Change the Social Cost of Selfish Routing?}
\author{Yu Cheng\footnote{Brown University. \href{mailto:yu_cheng@brown.edu}{\url{yu_cheng@brown.edu}}. Supported in part by NSF Award CCF-2307106.}
\and
Tianle Jiang\footnote{Duke University. \href{mailto:tianle.jiang@duke.edu}{\url{tianle.jiang@duke.edu}}. Supported in part by NSF Award IIS-2402823.}
}
\date{}

\begin{document}

\maketitle

\begin{abstract}
We study approximate Nash equilibria in nonatomic congestion games, with the goal of understanding how much the social cost can change when the equilibrium condition is relaxed.
We consider $\eps$-approximate Nash equilibria ($\eps$-NE), in which each player uses a strategy whose cost is at most $1 + \eps$ times the minimum available cost.
We compare the social costs of $\eps$-NE and exact Nash equilibria.
This lets us quantify, for example, how much a navigation app can reduce total travel time when drivers tolerate routes slightly slower than the fastest available route.

For degree-$d$ polynomial latencies in both general nonatomic congestion games and single-commodity routing games on series-parallel networks, we give tight lower and upper bounds on the relative change in social cost from an exact Nash equilibrium to any $\eps$-NE, with matching instances for every $\eps$.
For small $\eps$, the largest possible relative reduction and increase in general congestion games are both $\Theta(\sqrt{d\eps})$, independently of the number of resources and populations.
In contrast, on series-parallel networks of depth $h$, the largest possible relative reduction is $\Theta(h\eps)$, while the largest possible relative increase is known to be $\eps$.
For the classes of games we study, these results give a tight and quantitative characterization of the possible changes in social cost and establish a separation between general and series-parallel networks.
\end{abstract}

\section{Introduction}
\label{sec:intro}

Nonatomic routing games are a fundamental model in algorithmic game theory.
In this model, a continuum of infinitesimal drivers travels through a network, and the latency (travel time) of each edge depends on the total traffic using it.
A flow specifies how the traffic is split among the available routes.
A Nash equilibrium, also called a Nash flow or Wardrop equilibrium~\cite{wardrop1952}, is a flow in which every driver uses a route of minimum latency under the resulting traffic pattern.
The social cost of a flow is the total travel time of all drivers.
The classical examples of Pigou~\cite{pigou1920} and Braess~\cite{braess1968} show that a Nash flow need not minimize the social cost.
The seminal work on the Price of Anarchy (PoA)~\cite{KoutsoupiasP99, RoughgardenT02_poa, Roughgarden02_poa} quantifies this inefficiency as the worst-case ratio of the social cost of a Nash flow to the optimal social cost.

The exact Nash equilibrium condition, however, is often too restrictive as a model of drivers' behavior.
In many scenarios, drivers may lack accurate and timely estimates of travel times, perceive routes with slight travel time differences as equally good, or prefer a familiar route even if it is slightly slower.
Drivers may also be willing to follow a navigation app's recommendation as long as its estimated travel time is close to that of the fastest route.
Routing platforms could use this tolerance to steer traffic toward more desirable flows.
This motivates us to study $\eps$-approximate Nash equilibria ($\eps$-NE), in which every driver uses a route whose latency is at most $1+\eps$ times that of the fastest available route under the resulting traffic pattern.

Using a Nash flow as a benchmark for uncoordinated traffic, we compare the social cost of an $\eps$-NE with this benchmark to quantify the potential cost reduction enabled by the drivers' tolerance.
More generally, all Nash flows have the same social cost, whereas $\eps$-NE can have a range of social costs, both below and above that of a Nash flow.
This leads to the main question of this paper:

\begin{quote}
\emph{How much can approximate Nash equilibria improve or worsen the social cost relative to a Nash equilibrium, and how does the answer depend on the latency functions and the network structure?}
\end{quote}

\subsection{Problem Setup: Approximate Equilibria and Cost Ratios}

We first define approximate equilibria in the setting of single-commodity nonatomic routing games.
Our results for general networks also hold for nonatomic congestion games (defined in Section~\ref{sec:def_routing}).

A nonatomic routing instance consists of a directed network $G = (V, E)$ with source $s$ and sink~$t$.
Each edge $e \in E$ has a continuous, nonnegative, nondecreasing latency function $\ell_e$.
One unit of traffic must be routed from $s$ to $t$.
Let $\mathcal{P}$ denote the set of $s$-$t$ paths.
A flow $f$ assigns a nonnegative amount of traffic $f_P$ to each path $P \in \mathcal{P}$, with $\sum_{P \in \mathcal{P}} f_P = 1$.
Under $f$, an edge $e$ with total traffic~$x$ has latency $\ell_e(x)$, and the latency $\ell_P(f)$ of a path $P$ is the sum of its edge latencies.
The social cost is $C(f) = \sum_{P \in \mathcal{P}} f_P \ell_P(f)$.
A flow $f$ is a Nash flow if every used path has minimum latency, that is, $\ell_P(f) \le \ell_{P'}(f)$ for every path $P$ with $f_P > 0$ and every $P' \in \mathcal{P}$.

\begin{definition}[$\eps$-approximate Nash equilibrium]
A flow $f$ is an $\eps$-approximate Nash equilibrium ($\eps$-NE) if for every path $P$ with $f_P > 0$,
\[
\ell_P(f) \le (1 + \eps)\ell_{P'}(f)
\qquad \text{for all } P' \in \mathcal{P}.
\]
\end{definition}
All Nash flows of an instance have the same social cost.
We study how small or large the ratio $C(f')/C(f)$ can be as a function of $\eps$, where $f$ is a Nash flow and $f'$ is an $\eps$-NE.

\begin{example}
\label{ex:Pigou}
Consider a concatenation of $h \ge 2$ copies of Pigou's network as shown in Figure~\ref{fig:h-pigou-intro}.
Each copy has two parallel edges: an upper edge with latency $x$ and a lower edge with constant latency $1$.
The Nash flow $f$ routes all traffic along the all-upper path.
Every upper edge then has latency $1$, so the social cost is $C(f) = h$.

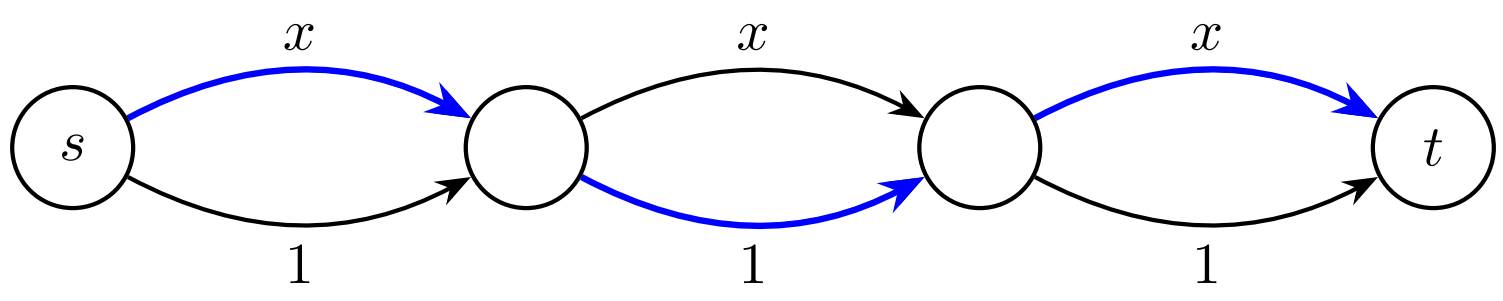
\begin{figure}[ht]
\centering
\begin{tikzpicture}[scale=1.0, auto, thick, >=Stealth]
    \node[node] (s) at (0,0) {$s$};
    \node[node] (v1) at (3,0) {};
    \node[node] (v2) at (6,0) {};
    \node[node] (t) at (9,0) {$t$};

    \draw[->,edge] (s) to[bend left=28] node[midway,above] {$x$} (v1);
    \draw[->,edge] (s) to[bend right=28] node[midway,below] {$1$} (v1);

    \draw[->,edge] (v1) to[bend left=28] node[midway,above] {$x$} (v2);
    \draw[->,edge] (v1) to[bend right=28] node[midway,below] {$1$} (v2);

    \draw[->,edge] (v2) to[bend left=28] node[midway,above] {$x$} (t);
    \draw[->,edge] (v2) to[bend right=28] node[midway,below] {$1$} (t);

    \draw[->,very thick,blue] (s) to[bend left=28] (v1);
    \draw[->,very thick,blue] (v1) to[bend right=28] (v2);
    \draw[->,very thick,blue] (v2) to[bend left=28] (t);
\end{tikzpicture}
\caption{A concatenation of $h$ copies of Pigou's network for $h = 3$.
The blue path is one of the paths used by the $\eps$-NE $f'$ described below.}
\label{fig:h-pigou-intro}
\end{figure}

Consider the flow $f'$ that routes $\frac{1}{h}$ units of traffic on each of the $h$ paths that use exactly one lower edge.
Every upper edge then carries $1 - \frac{1}{h}$ units of traffic, so the all-upper path is a shortest path with latency $h - 1$.
Every used path has latency $h - 1 + \frac{1}{h}$.
Thus $f'$ is an $\eps$-NE with
\[
\eps = \frac{1}{h(h-1)},
\qquad
\frac{C(f')}{C(f)} = \frac{h - 1 + \frac{1}{h}}{h} = 1 - \frac{1}{h} + \frac{1}{h^2}.
\]
For $h = 10$, every driver's route is only about $1\%$ slower than the fastest under the new traffic pattern, yet their travel time is $9\%$ lower than in the Nash flow.
More generally, this construction gives a relative reduction in social cost of $\Theta(1/h) = \Theta(\sqrt{\eps})$.
In contrast, on a single copy of Pigou's network, the largest possible relative reduction is only $\Theta(\eps)$.
\end{example}

\paragraph{Cost ratios.}
For a class $\mathcal{G}$ of nonatomic congestion games and a class $\mathcal{L}$ of latency functions, let $\mathcal{I}(\mathcal{G},\mathcal{L})$ denote the instances in $\mathcal{G}$ with latency functions in $\mathcal{L}$.
For example, $\mathcal{G}$ may be all nonatomic congestion games or routing games on series-parallel networks, and $\mathcal{L}$ the degree-$d$ polynomials.
We~define
\begin{align*}
\best(\mathcal{G},\mathcal{L}) &= \min_{I \in \mathcal{I}(\mathcal{G},\mathcal{L}),\ \text{$\eps$-NE $f'$ of $I$}} \frac{C(f')}{C(f)},\\
\worst(\mathcal{G},\mathcal{L}) &= \max_{I \in \mathcal{I}(\mathcal{G},\mathcal{L}),\ \text{$\eps$-NE $f'$ of $I$}} \frac{C(f')}{C(f)},
\end{align*}
where $f$ is a Nash equilibrium of $I$.
A lower bound on $\best$ limits the relative reduction in social cost, while an upper bound on $\worst$ limits the relative increase.

\subsection{Our Results}

Our main results give a tight and quantitative characterization of how much approximate equilibria can change the social cost, and how this depends on the latency functions and the network structure.
For degree-$d$ polynomial latencies, Table~\ref{tab:overview} summarizes our lower and upper bounds on $C(f')/C(f)$ for general nonatomic congestion games and for single-commodity routing on series-parallel networks.

\begin{table}[ht]
\centering
\small
\renewcommand{\arraystretch}{1.2}
\begin{tabular}{|c|l|l|l|}
\hline
Setting & Bound on $C(f')/C(f)$ & Proof & Matching instance \\
\hline
\multirow{2}{*}{{\renewcommand{\arraystretch}{0.9}\begin{tabular}{@{}c@{}}General nonatomic\\congestion games\end{tabular}}} & $\ge 1 - \sqrt{d\eps} + O_d(\eps)$ & Corollary~\ref{cor:general_polynomial_lowerbound} & Example~\ref{ex:general_polynomial_lowerbound_tight} \\
\cline{2-4}
& $\le 1 + \sqrt{d\eps} + O_d(\eps)$ & Theorem~\ref{thm:general_polynomial_upperbound} & Example~\ref{ex:general_polynomial_upperbound_tight} \\
\hline
\multirow{2}{*}{{\renewcommand{\arraystretch}{0.9}\begin{tabular}{@{}c@{}}Nonatomic routing games on\\depth-$h$ series-parallel networks\end{tabular}}} & $\ge 1 - h\eps + O_{d,h}(\eps^2)$ & Corollary~\ref{cor:series_parallel_une_polynomial_lowerbound} & Example~\ref{ex:h-pigou_polynomial} \\
\cline{2-4}
& $\le 1 + \eps$~\cite{sagt/KleerS17} & Theorem~\ref{thm:series_parallel_upperbound} & Example~\ref{ex:sp-upper-tight} \\
\hline
\end{tabular}
\caption{Summary of our tight bounds on $C(f')/C(f)$ for degree-$d$ polynomial latencies, shown to leading order in $\eps$, where $f$ is a Nash equilibrium and $f'$ is an $\eps$-NE.\protect\footnotemark{}
Sections~\ref{sec:best} and~\ref{sec:worst} give the exact bounds for every $\eps$.}
\label{tab:overview}
\end{table}
\footnotetext{Throughout the paper, the constants in $O_d(\cdot)$ and $O_{d,h}(\cdot)$ depend only on the subscripted parameters.}

For general nonatomic congestion games, which allow multiple populations with different demands, we show that the relative change in social cost from a Nash equilibrium to any $\eps$-NE is at most $\sqrt{d\eps} + O_d(\eps)$ in either direction, independently of the number of resources and populations.
Our exact lower and upper bounds are tight for every $\eps$, including all lower-order terms.

\begin{figure}[ht]
\centering
\begin{minipage}[b]{0.49\textwidth}
\centering
\begin{tikzpicture}[x=11.2cm,y=3.2cm,font=\small]
    \begin{scope}
        \clip (0,0.575) rectangle (0.55,2.3);
        \fill[blue!10]
            plot[domain=0:0.57735,samples=80] ({\x*\x},{1/(1-\x/sqrt(1+\x*\x))})
            -- plot[domain=1/3:0.55,samples=40] ({\x},{4*(1+\x)/(3-\x)})
            -- (0.55,0.75) -- (0.5,0.75)
            -- plot[domain=0.5:0,samples=80] ({\x*\x/(1-\x)},{1-\x+\x*\x})
            -- cycle;
        \draw[thin,gray] (0,1) -- (0.55,1);
        \draw[very thick,blue]
            plot[domain=0:0.57735,samples=80] ({\x*\x},{1/(1-\x/sqrt(1+\x*\x))})
            -- plot[domain=1/3:0.55,samples=40] ({\x},{4*(1+\x)/(3-\x)});
        \draw[very thick,blue]
            plot[domain=0:0.5,samples=80] ({\x*\x/(1-\x)},{1-\x+\x*\x})
            -- (0.55,0.75);
        \draw[thick,dashed]
            plot[domain=0:0.55,samples=60] ({\x},{4*(1+\x)/(3-\x)});
        \draw[thick,dashed] (0,0.75) -- (0.55,0.75);
    \end{scope}

    \draw[gray,dotted] ({1/3},0.575) -- ({1/3},2);
    \draw[gray,dotted] (0.5,0.575) -- (0.5,0.75);
    \fill ({1/3},2) circle (1.5pt);
    \fill (0.5,0.75) circle (1.5pt);

    \draw[->] (0,0.575) -- (0.58,0.575) node[right] {$\eps$};
    \draw[->] (0,0.575) -- (0,2.4) node[above] {$C(f')/C(f)$};
    \foreach \y/\lab in {0.75/\frac{3}{4}, 1/1, 1.3333/\frac{4}{3}, 2/2}
        \draw (0,\y) -- (-0.008,\y) node[left] {$\lab$};
    \foreach \x/\lab in {0/0, 0.3333/\frac{1}{3}, 0.5/\frac{1}{2}}
        \draw (\x,0.575) -- (\x,0.545) node[below] {$\lab$};

    \node[rotate=29,anchor=south,inner sep=3pt] at (0.15,1.614) {approximate PoA};
    \node[rotate=0,anchor=north,inner sep=3pt] at (0.3,0.750) {$1/\mathrm{PoA}$};
    \node[blue,rotate=34,anchor=north,inner sep=3pt] at (0.22,1.738) {$\worst$};
    \node[blue,rotate=-3,anchor=south,inner sep=3pt] at (0.22,0.766) {$\best$};
\end{tikzpicture}
\subcaption{Our bounds versus PoA-based bounds}
\end{minipage}
\hfill
\begin{minipage}[b]{0.49\textwidth}
\centering
\begin{tikzpicture}[x=11.2cm,y=3.2cm,font=\small]
    \begin{scope}
        \clip (0,0.575) rectangle (0.55,2.3);
        \fill[blue!6]
            plot[domain=0:0.57735,samples=80] ({\x*\x},{1/(1-\x/sqrt(1+\x*\x))})
            -- plot[domain=1/3:0.55,samples=40] ({\x},{4*(1+\x)/(3-\x)})
            -- (0.55,0.75) -- (0.5,0.75)
            -- plot[domain=0.5:0,samples=80] ({\x*\x/(1-\x)},{1-\x+\x*\x})
            -- cycle;
        \fill[red!10]
            (0,1) -- (0.55,1.55) -- (0.55,0.75) -- (0.5,0.75)
            -- plot[domain=0.5:1/3,samples=40] ({\x*\x/(1-\x)},{1-\x+\x*\x})
            -- plot[domain=1/3:0,samples=60] ({\x/(3*(1-\x))},{1-\x+\x*\x})
            -- cycle;
        \draw[thin,gray] (0,1) -- (0.55,1);
        \draw[thick,blue]
            plot[domain=0:0.57735,samples=80] ({\x*\x},{1/(1-\x/sqrt(1+\x*\x))})
            -- plot[domain=1/3:0.55,samples=40] ({\x},{4*(1+\x)/(3-\x)});
        \draw[thick,blue]
            plot[domain=0:0.5,samples=80] ({\x*\x/(1-\x)},{1-\x+\x*\x})
            -- (0.55,0.75);
        \draw[very thick,red] (0,1) -- (0.55,1.55);
        \draw[very thick,red]
            plot[domain=0:1/3,samples=60] ({\x/(3*(1-\x))},{1-\x+\x*\x})
            -- plot[domain=1/3:0.5,samples=40] ({\x*\x/(1-\x)},{1-\x+\x*\x})
            -- (0.55,0.75);
        \draw[very thick,red,dotted]
            plot[domain=0:0.35484,samples=60] ({\x/(1-\x)},{1-\x+\x*\x});
    \end{scope}

    \draw[gray,dotted] ({1/6},0.575) -- ({1/6},{7/9});
    \draw[gray,dotted] (0.5,0.575) -- (0.5,0.75);
    \fill ({1/6},{7/9}) circle (1.5pt);
    \fill (0.5,0.75) circle (1.5pt);

    \draw[->] (0,0.575) -- (0.58,0.575) node[right] {$\eps$};
    \draw[->] (0,0.575) -- (0,2.4) node[above] {$C(f')/C(f)$};
    \foreach \y/\lab in {0.75/\frac{3}{4}, 1/1, 2/2}
        \draw (0,\y) -- (-0.008,\y) node[left] {$\lab$};
    \foreach \x/\lab in {0/0, 0.16667/\frac{1}{6}, 0.5/\frac{1}{2}}
        \draw (\x,0.575) -- (\x,0.545) node[below] {$\lab$};

    \node[blue,rotate=37,anchor=south,inner sep=3pt] at (0.14,1.539) {general};
    \node[red,rotate=16,anchor=south,inner sep=3pt] at (0.3,1.300) {series-parallel};
    \node[red,rotate=-3,anchor=south,inner sep=3pt] at (0.42,0.792) {$h = 1$};
    \node[red,rotate=-1,anchor=north,inner sep=3pt] at (0.3,0.757) {$h = 3$};
\end{tikzpicture}
\subcaption{General versus series-parallel networks}
\end{minipage}
\caption{Bounds on $C(f')/C(f)$ for linear latencies, where $f$ is a Nash equilibrium and $f'$ is an $\eps$-NE.
(a) Our tight bounds for general congestion games (solid) and the bounds implied by the PoA and the approximate PoA (dashed).
(b) Bounds for general congestion games and series-parallel networks.
The series-parallel lower bound is shown for depths $h = 1$ (dotted) and $h = 3$, and the upper bound $1 + \eps$ holds for every depth.}
\label{fig:band}
\end{figure}
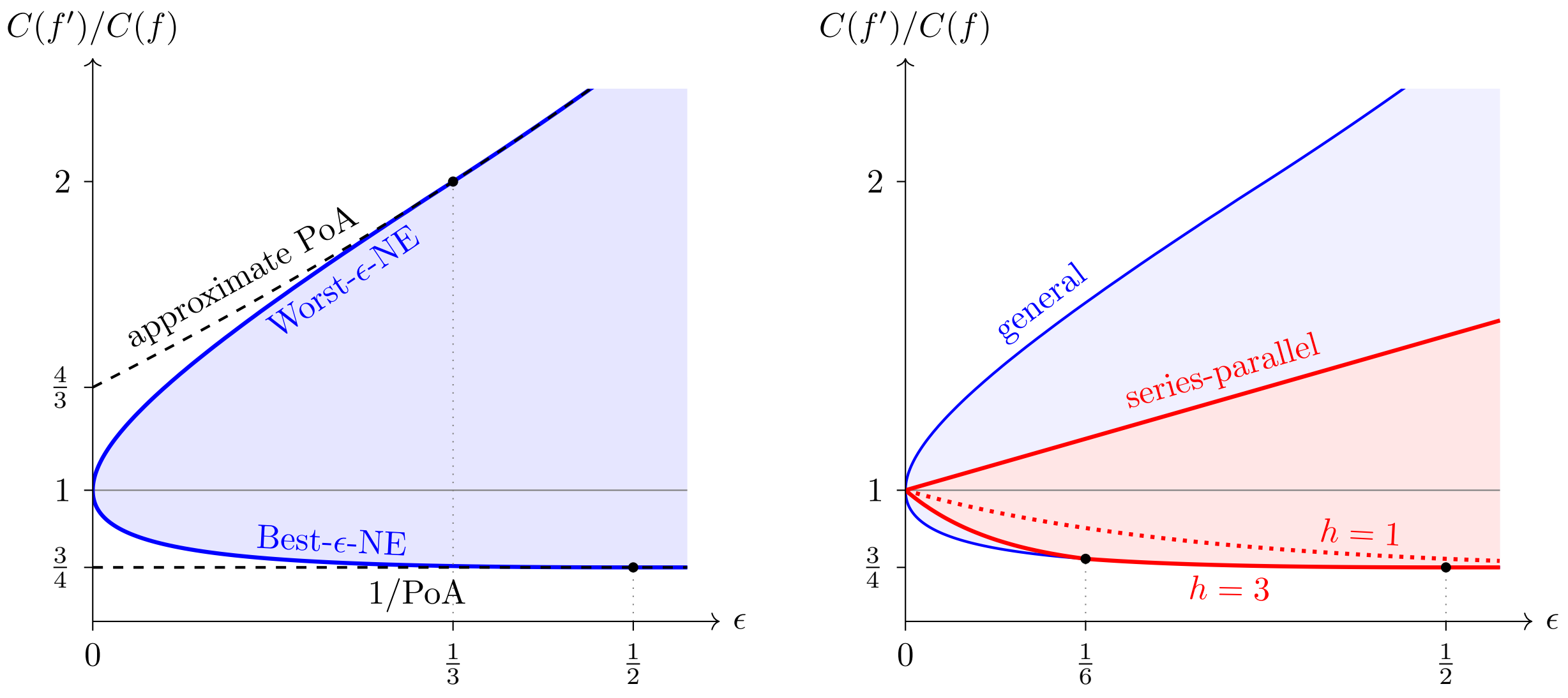

For single-commodity routing games on series-parallel networks of depth $h$ (the maximum number of edges on an $s$-$t$ path), we show that the largest possible relative reduction in social cost is $h\eps + O_{d,h}(\eps^2)$, so the degree $d$ affects only lower-order terms (Figure~\ref{fig:band}(b)).
The largest possible relative increase is $\eps$~\cite{sagt/KleerS17}.
The separation between $\sqrt{\eps}$ and $\eps$ comes from network topology rather than from path length.
We construct instances on Braess graphs in which the relative change is of order $\sqrt{\eps}$ in both directions, even though every $s$-$t$ path has at most three edges (Appendix~\ref{sec:braess_tight_examples}).
Our results extend to arbitrary monotone latencies, for which we give tight bounds that separate general networks from series-parallel ones (Section~\ref{sec:monotone_lower}).

\paragraph{Comparison with PoA and PoS.}
The approximate versions of the PoA and the Price of Stability (PoS) compare the costs of the worst and best $\eps$-NE, respectively, with the minimum social cost $\mathrm{OPT}$~\cite{ChristodoulouKS11}.
The classical PoA bounds $C(f)/\mathrm{OPT}$, and the approximate PoA bounds $C(f')/\mathrm{OPT}$.
Since $\mathrm{OPT}$ is at most both $C(f)$ and $C(f')$, these bounds imply that for linear latencies and $\eps \le 1$,
\[
\frac{3}{4}
\le \frac{\mathrm{OPT}}{C(f)}
\le \frac{C(f')}{C(f)}
\le \frac{C(f')}{\mathrm{OPT}}
\le \frac{4(1+\eps)}{3-\eps}.
\]

As $\eps \to 0$, these bounds approach $\frac{3}{4}$ and $\frac{4}{3}$, whereas ours converge to $1$ (Figure~\ref{fig:band}(a)).
The same contrast holds for every fixed polynomial degree $d$.
For linear latencies, our upper and lower bounds coincide with the PoA-based bounds for $\eps \ge \frac{1}{3}$ and $\eps \ge \frac{1}{2}$, respectively.

Moreover, our bounds show that the largest possible relative changes in social cost are of order $\eps$ on series-parallel networks of fixed depth but $\sqrt{\eps}$ on general networks (Figure~\ref{fig:band}(b)).
In contrast, the classical PoA bound for degree-$d$ polynomial latencies is already tight on two parallel edges~\cite{Roughgarden02_poa}, so it is the same for both classes.

Approximate PoS gives no bound on $C(f')/C(f)$ that holds for every $\eps$-NE $f'$.
It only guarantees that some $\eps$-NE has cost at most a given factor times $\mathrm{OPT}$.

\subsection{Technical Overview}

Let $f$ be a Nash equilibrium and $f'$ an $\eps$-NE of a nonatomic congestion game.
As is common in classical PoA analyses~\cite{Roughgarden02_poa, RoughgardenT04, geb/CorreaSM08}, we define $C_{f'}(f)$ as the cost of $f$ at the latencies induced by $f'$, and define $C_f(f')$ analogously.
By Lemma~\ref{lem:residual_inequality}, the equilibrium condition for $f$ implies $C(f)\le C_f(f')$, and the approximate-equilibrium condition for $f'$ implies $C(f')\le(1+\eps)C_{f'}(f)$.

\paragraph{The lower bounds on $\best$.}
Our lower bounds for general nonatomic congestion games and for single-commodity routing on series-parallel networks follow from a single theorem (Theorem~\ref{thm:general_polynomial_meta_lowerbound}).
The proof compares $C(f)$ and $C(f')$ through $C_{f'}(f)$:
\[
C(f)-C(f')
= \underbrace{\bigl(C(f)-C_{f'}(f)\bigr)}_{T}
- \underbrace{\bigl(C(f')-C_{f'}(f)\bigr)}_{R}.
\]
The term $T$ is the reduction in the cost of $f$ when evaluated at the latencies induced by $f'$.
The term $R$ is the additional cost of $f'$ over $f$ when both are evaluated at the latencies induced by $f'$.

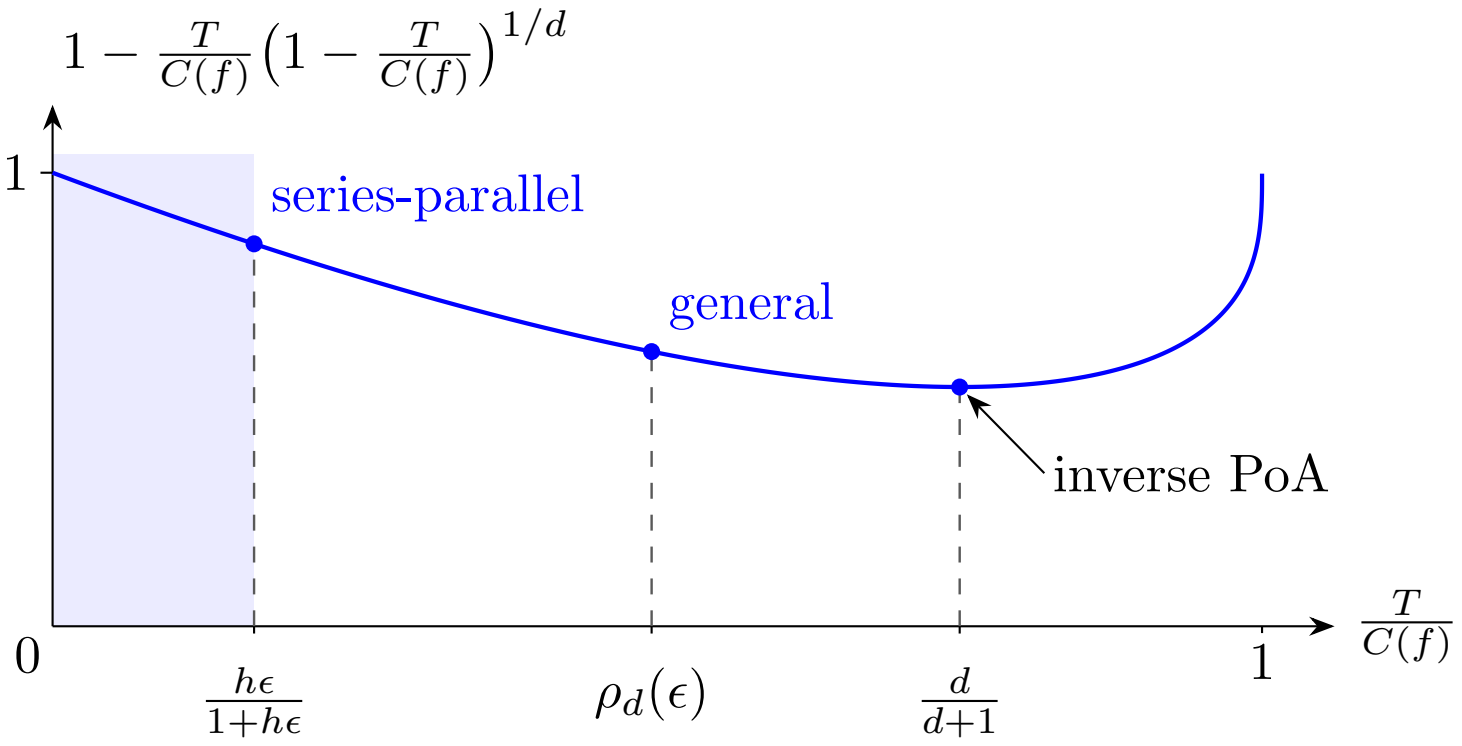
\begin{figure}[ht]
\centering
\begin{tikzpicture}[x=8cm,y=3cm,>=Stealth,font=\small]
    \def\dvalue{3}
    \def\hvalue{1}
    \def\epsvalue{0.2}
    \pgfmathsetmacro{\networkbound}{\hvalue*\epsvalue/(1+\hvalue*\epsvalue)}
    \def\rhobound{0.495281643544}
    \pgfmathsetmacro{\poapoint}{\dvalue/(\dvalue+1)}
    \pgfmathsetmacro{\networkheight}{1-\networkbound*pow(1-\networkbound,1/\dvalue)}
    \pgfmathsetmacro{\rhoheight}{1-\rhobound*pow(1-\rhobound,1/\dvalue)}
    \pgfmathsetmacro{\poaheight}{1-\poapoint*pow(1-\poapoint,1/\dvalue)}
    \fill[blue!8] (0,0) rectangle (\networkbound,1.04);
    \draw[->] (0,0) -- (1.06,0) node[right] {$\frac{T}{C(f)}$};
    \draw[->] (0,0) -- (0,1.15) node[above right,inner sep=1.5pt]
        {$1-\tfrac{T}{C(f)}\bigl(1-\tfrac{T}{C(f)}\bigr)^{1/d}$};
    \node[below left,inner sep=2pt] at (0,0) {$0$};
    \draw (0,1) -- (-0.01,1) node[left,inner sep=2pt] {$1$};
    \draw (1,0) -- (1,-0.015) node[below,inner sep=2pt] {$1$};
    \draw[dashed,black!65,line width=.45pt] (\networkbound,0) -- (\networkbound,\networkheight);
    \draw[dashed,black!65,line width=.45pt] (\rhobound,0) -- (\rhobound,\rhoheight);
    \draw[dashed,black!65,line width=.45pt] (\poapoint,0) -- (\poapoint,\poaheight);
    \draw[thick,blue,domain=1:0,samples=151,variable=\u]
        plot ({1-pow(\u,\dvalue)},{1-\u+pow(\u,\dvalue+1)});
    \fill[blue] (\networkbound,\networkheight) circle (1.6pt);
    \fill[blue] (\rhobound,\rhoheight) circle (1.6pt);
    \fill[blue] (\poapoint,\poaheight) circle (1.6pt);
    \node[blue,anchor=south west,inner sep=2pt,xshift=1pt,yshift=2pt]
        at (\networkbound,\networkheight) {series-parallel};
    \node[blue,anchor=south west,inner sep=2pt,xshift=1pt,yshift=2pt]
        at (\rhobound,\rhoheight) {general};
    \draw (\networkbound,0) -- (\networkbound,-.016);
    \node[anchor=north,inner sep=1.5pt] at (\networkbound,-.071) {$\frac{h\eps}{1+h\eps}$};
    \draw (\rhobound,0) -- (\rhobound,-.016);
    \node[anchor=north,inner sep=1.5pt] at (\rhobound,-.071) {$\rho_d(\eps)$};
    \draw (\poapoint,0) -- (\poapoint,-.016);
    \node[anchor=north,inner sep=1.5pt] at (\poapoint,-.071) {$\frac{d}{d+1}$};
    \node[anchor=west,inner sep=1.4pt] (poa) at (0.82,0.33753) {inverse PoA};
    \draw[->,shorten >=1.9pt] (poa.west) -- (\poapoint,\poaheight);
\end{tikzpicture}
\caption{Illustration of Theorem~\ref{thm:general_polynomial_meta_lowerbound} for polynomial latency functions of degree $d=3$ and $\eps=0.2$.
The points labeled \emph{general} and \emph{series-parallel} give the lower bounds on $C(f')/C(f)$ for general congestion games and for routing games on series-parallel networks of depth $h=1$, respectively.
The shaded interval contains the possible values of $T/C(f)$ for the series-parallel case.}
\label{fig:unified-lower-bound}
\end{figure}

Intuitively, to minimize $C(f')$, we want to maximize $T$ and minimize $R$.
We first lower bound $R$ in terms of $T$ and $d$, where $d$ is the maximum degree of the polynomial latency functions.
If the load on a resource $e$ falls from $x$ to $qx$, where $0\le q\le1$, its polynomial latency satisfies $\ell_e(qx)\ge q^d\ell_e(x)$.
Combining this inequality with $C_f(f')\ge C(f)$ and Jensen's inequality yields
\[
R \ge T\left[1-\left(1-\tfrac{T}{C(f)}\right)^{1/d}\right],
\qquad\text{and hence}\qquad
\frac{C(f')}{C(f)}
\ge 1-\tfrac{T}{C(f)}\left(1-\tfrac{T}{C(f)}\right)^{1/d}.
\]
As a function of $\tfrac{T}{C(f)}$, the right-hand side decreases up to $\tfrac{d}{d+1}$ and then increases (Figure~\ref{fig:unified-lower-bound}).

It remains to decide how large $\tfrac{T}{C(f)}$ can be.
Since $C(f')\le(1+\eps)C_{f'}(f)$, we have $R\le\eps(C(f)-T)$.
We define $\rho_d(\eps)$ as the largest value of $\tfrac{T}{C(f)}$ for which the earlier lower bound on $R$ does not exceed $\eps(C(f)-T)$.
Thus $\tfrac{T}{C(f)}\le\rho_d(\eps)$, where $\rho_d(\eps) \approx \sqrt{d\eps}$ for small $\eps$.
Our unified approach reduces the task of proving a sharper lower bound for a more structured class of games to a single step: narrowing the range of $\tfrac{T}{C(f)}$.
For example, for series-parallel networks of depth $h$, we prove that $C_{f'}(f)\ge\tfrac{C(f)}{1+h\eps}$ and hence $\tfrac{T}{C(f)}\le\tfrac{h\eps}{1+h\eps}$ (Lemma~\ref{lem:series_parallel_usedpath_lowerbound}).

At $\tfrac{T}{C(f)}=\tfrac{d}{d+1}$, the curve in Figure~\ref{fig:unified-lower-bound} reaches its minimum, which equals the inverse of the classical PoA bound (Remark~\ref{remark:poa}).
For small $\eps$, however, the range of $\tfrac{T}{C(f)}$ ends before $\tfrac{d}{d+1}$, so the minimum over this range occurs at its right endpoint.
For general congestion games, evaluating the curve at $\rho_d(\eps)$ gives a lower bound of $1-\sqrt{d\eps}+O_d(\eps)$.
On series-parallel networks of fixed depth $h$, evaluating the curve at $\tfrac{h\eps}{1+h\eps}$ gives a lower bound of $1-h\eps+O_{d,h}(\eps^2)$.

Our matching instances show that our lower bounds are tight for every $\eps$.
Our proof identifies $T$ as the key quantity that determines how much an approximate equilibrium can be cheaper than a Nash equilibrium.
The equilibrium condition for $f$ and the latency functions tell us how much of a given reduction $T$ must be offset by the additional cost $R$, while the approximate-equilibrium condition for $f'$ and the network structure limit how large that reduction can be.

\paragraph{The upper bound on $\worst$.}
Our tight upper bound has three regimes of $\eps$ (Theorem~\ref{thm:general_polynomial_upperbound}), and our proof uses a unified primal-dual framework to derive all three.
As in the lower-bound proof, we compare two bounds on the additional cost $R$.
We fix a candidate value $\alpha$ for the cost ratio $\tfrac{C(f')}{C(f)}$ and ask how small $R$ can be at that ratio.
Our primal optimization problem minimizes $R$ over resource loads for $f$ and $f'$ subject to $C(f')=\alpha C(f)$ and the equilibrium condition $C_f(f')\ge C(f)$.

We analyze the Lagrangian dual of this problem.
Each feasible dual solution gives a lower bound on $\tfrac{R}{C(f)}$.
On the other hand, the approximate-equilibrium condition $C(f')\le(1+\eps)C_{f'}(f)$ gives $\tfrac{R}{C(f)}\le\tfrac{\eps}{1+\eps}\alpha$.
If the lower bound exceeds this value, then no $\eps$-NE can attain the cost ratio $\alpha$.

The key dual variable is the multiplier $A$ for the equilibrium condition $C_f(f')\ge C(f)$.
Setting $A=0$ gives a bound without using the fact that $f$ is an equilibrium.
This recovers the approximate PoA bounds of Christodoulou et~al.~\cite{ChristodoulouKS11}, which remain tight for $\tfrac{C(f')}{C(f)}$ in the second and third regimes.
In the first regime, where $\eps$ is small, giving the equilibrium condition positive weight yields the sharper bound, which is $1+\sqrt{d\eps}+O_d(\eps)$.
Therefore, the equilibrium condition for $f$ strengthens the bound only for small $\eps$, precisely where $\worst$ is smaller than the approximate PoA.

\subsection{Related Work}

\paragraph{PoA and PoS.}
Tight PoA bounds for nonatomic routing and congestion games were established in the seminal works of Roughgarden and Tardos~\cite{RoughgardenT02_poa, RoughgardenT04} and Roughgarden~\cite{Roughgarden02_poa}; see also~\cite{geb/CorreaSM08}.
In atomic congestion games, equilibria can have different social costs, and both the PoA and the PoS have been studied extensively~\cite{AwerbuchAE05, ChristodoulouK05_1, ChristodoulouK05_2, AlandDGMS06, BhawalkarGR10, ChristodoulouG13, 0001GGS18}.
The PoA and PoS have also been studied in other fundamental problems in algorithmic game theory, such as network design~\cite{AnshelevichDKTWR04, BiloFM13} and auctions~\cite{LucierB10, LemeT10, SyrgkanisT13, HartlineHT14, RoughgardenST17, stoc/HoyTW18, JinL22, JinL23}.

Primal-dual methods have been a useful technique especially for proving robust PoA~\cite{NadavR10, Bilo12, KulkarniM15}, which quantifies the inefficiency of equilibrium concepts that are weaker than Nash equilibrium, such as correlated equilibrium and coarse correlated equilibrium.
We refer the reader to Bil{\`{o}}~\cite{Bilo12} for a more comprehensive review of the application of the primal-dual method to various other settings.
Unlike prior formulations in which the primal program directly optimizes the PoA or social cost of the target equilibrium concept, our primal program minimizes an intermediate quantity at a fixed cost ratio.

\paragraph{Approximate equilibria and related notions.}
Christodoulou et~al.~\cite{ChristodoulouKS11} study approximate PoA and PoS in congestion games.
Basu et~al.~\cite{BasuYLNC17} study approximate equilibria and related fairness notions in nonatomic routing, including the computational complexity of finding minimum-cost flows satisfying these constraints.
Transportation models of boundedly rational user equilibrium (BRUE) allow drivers to use routes within an indifference band of the shortest-path cost.
Lou et~al.~\cite{LouYL10} study best- and worst-cost BRUE flows and robust congestion pricing, while Di et~al.~\cite{DiLPB13,DiHGL14} study the equilibrium solution sets and the Braess paradox in this setting.

Kleer and Sch\"afer~\cite{sagt/KleerS17} study the worst-case ratio between the social costs of an $\eps$-approximate Nash flow and an exact Nash flow, which they call the \emph{$\eps$-stability ratio}.
For single-commodity routing with arbitrary monotone latencies, they derive upper bounds that depend on the number of vertices.
Takalloo and Kwon~\cite{TakallooK20} study the same cost comparison under the name \emph{price of satisficing}.
For polynomial latencies, their PoA-based upper bounds coincide with ours in the second and third regimes.
For small $\eps$, they obtain the bound $(1+\eps)^{d+1}$ under an additional condition.
Our lower and upper bounds for general congestion games with polynomial latencies depend only on $d$ and $\eps$ and require no such condition.
Nikolova and Stier-Moses~\cite{NikolovaS15} study the \emph{price of risk aversion}, which compares the social costs of risk-averse and risk-neutral equilibria.
Kleer and Sch\"afer~\cite{sagt/KleerS16} study the \emph{deviation ratio}, which compares equilibria under perturbed and original latencies, with social cost measured using the original latencies.

\paragraph{Other variants of nonatomic routing games.}
There is a broad literature that studies variants of the classical nonatomic routing game, with the aim of understanding how the social cost of selfish routing can change when users' behavior, the network itself, the information available to users, or the traffic demand is changed.
In Stackelberg routing or routing with altruism~\cite{Roughgarden01_stackelberg, KumarM02, ChenK08, SharmaW09, Swamy12, ChenKKS14}, part of the traffic is either controlled by a central planner or is not routed in a completely selfish manner, while the remaining traffic routes selfishly.
Other work~\cite{Roughgarden01_design, ColeDR03, ColeDR06, LinRT04, LinRTW05, BhaskarLSS14} considers interventions on users' incentives or feasible routes by imposing taxes or tolls on the edges, or even removing some edges, in order to avoid the Braess paradox and improve the social cost at equilibrium.
A different approach is to reduce social cost at equilibrium by choosing what information about uncertain traffic conditions to reveal to drivers~\cite{BhaskarCKS16,Dughmi17,Kamenica19,GriesbachHKK22}.
Another line of work~\cite{EnglertFO08, WangDC13, CominettiDS19, Colini-Baldeschi20, CominettiDS24} examines how equilibrium outcomes vary with the traffic demand.

\paragraph{Series-parallel networks and the Braess paradox.}
Series-parallel networks form a canonical special class in routing games.
Milchtaich~\cite{geb/Milchtaich06} shows that the Braess paradox~\cite{braess1968} can occur in an undirected two-terminal network if and only if it is not series-parallel.
Subsequent work establishes that in various settings, series-parallel networks have benign equilibrium behavior~\cite{geb/EpsteinFM09a, CominettiDS19, HillB24, NikolovaS15, geb/MacaultST22}, or admit efficient algorithms~\cite{Swamy12, ipco/BhaskarLS14, BhaskarLSS14, Sperber10}.
In contrast, general networks containing Braess-like structures have long served as a source of inefficiency and computational hardness, for example in the Stackelberg routing model and in the problem of computing taxes~\cite{wine/BonifaciHS08, Roughgarden01_design, LinRT04, LinRTW05}.
The classical family of Braess graphs has been used to demonstrate several extremal phenomena~\cite{Roughgarden01_design, ColeDR06, EnglertFO08, sagt/KleerS17}.
We construct instances on this family that give essentially tight bounds for $\best$ and $\worst$ in our model.
Our results add to this distinction between series-parallel networks and general networks: the social cost of approximate equilibria is more stable on series-parallel networks than on general networks.

\section{Preliminaries}
\label{sec:prelim}

\subsection{Nonatomic Congestion Games}
\label{sec:def_routing}

A nonatomic congestion game consists of a finite set of resources $E$ and a finite set of populations $K$.
Each population $i \in K$ has demand $r_i > 0$ and a nonempty strategy set $\mathcal{S}_i \subseteq 2^E$.
Each resource $e \in E$ has a continuous, nonnegative, nondecreasing latency function $\ell_e:\mathbb{R}_{\ge 0} \to \mathbb{R}_{\ge 0}$.
A flow $f = (f_{i,S})_{i\in K,S\in\mathcal{S}_i}$ specifies the amount of population $i$ using each strategy $S \in \mathcal{S}_i$, where $f_{i,S} \ge 0$ and $\sum_{S\in\mathcal{S}_i}f_{i,S} = r_i$ for every $i \in K$.
The induced load on resource $e$ is
\[
x_e = \sum_{i\in K}\sum_{S\in\mathcal{S}_i:e\in S}f_{i,S}.
\]
The latency of a strategy $S \in \mathcal{S}_i$ is $\ell_S(f) = \sum_{e\in S}\ell_e(x_e)$, and the social cost is
\[
C(f) = \sum_{i\in K}\sum_{S\in\mathcal{S}_i}f_{i,S}\ell_S(f) = \sum_{e\in E}x_e\ell_e(x_e).
\]
We refer to a flow that minimizes the social cost as a \emph{socially optimal flow}.

A Nash equilibrium is a flow $f$ in which every used strategy has minimum latency: for every population $i \in K$ and strategy $S \in \mathcal{S}_i$ with $f_{i,S} > 0$, we have
\[
\ell_S(f) \le \ell_{S'}(f) \qquad \text{for all } S' \in \mathcal{S}_i.
\]
For $\eps \ge 0$, a flow $f$ is an $\eps$-approximate Nash equilibrium ($\eps$-NE) if every used strategy has latency at most $1 + \eps$ times the minimum latency: for every population $i \in K$ and strategy $S \in \mathcal{S}_i$ with $f_{i,S} > 0$, we have
\[
\ell_S(f) \le (1 + \eps)\ell_{S'}(f) \qquad \text{for all } S' \in \mathcal{S}_i.
\]
It is well known that a Nash equilibrium exists and that all Nash equilibria have the same social cost.
Given a Nash equilibrium $f$ and an $\eps$-NE $f'$, we compare their social costs through the ratio $C(f')/C(f)$.
Throughout the paper, we restrict attention to nondegenerate instances with $C(f) > 0$.

A nonatomic routing game is a special case of a nonatomic congestion game where the resources are the edges of a directed network, each population is usually referred to as a commodity with source $s_i$ and sink $t_i$, and the strategies of each commodity are the set of all $s_i$-$t_i$ paths.
In particular, a single-commodity routing game has one population whose strategies are the $s$-$t$ paths.

In this paper, we mainly consider three classes of latency functions:
\begin{itemize}
    \item Linear: $\ell_e(x) = a_ex + b_e$, where $a_e \ge 0, b_e \ge 0$.
    \item Polynomial of degree at most $d \ge 1$: $\ell_e(x) = \sum_{j=0}^d a_{e,j}x^j$, where $a_{e,j} \ge 0$ for all $0 \le j \le d$.
    \item Monotone: all continuous, nonnegative, nondecreasing functions.
\end{itemize}

\subsection{Useful Properties of Approximate Equilibria}

We first establish some basic notation and properties of approximate equilibria in nonatomic congestion games, which will be extensively used in our proofs.

\begin{definition}
Given two flows $g, g'$, let $C_{g}(g')$ denote the social cost of $g'$ when resource latencies are evaluated at the loads induced by $g$.
Formally, if $y$ and $z$ are the induced resource-load vectors of $g$ and $g'$, respectively, then
\[
\textstyle C_g(g') = \sum_{e\in E}z_e \ell_e(y_e).
\]
\end{definition}

\begin{lemma}[Folklore]
\label{lem:residual_inequality}
In any nonatomic congestion game, let $f$ and $f'$ be flows with induced resource-load vectors $x$ and $x'$, respectively.
\begin{enumerate}
    \item If $f$ is a Nash equilibrium, then $C_f(f') - C(f) \ge 0$.
    \item If $f'$ is an $\eps$-NE, then $C(f') \le (1 + \eps)C_{f'}(f)$.
    \item If $f$ is a Nash equilibrium and $f'$ is an $\eps$-NE, then $\sum_{e\in E} (x'_e - x_e)\cdot (\ell_e(x'_e) - \ell_e(x_e)) \le \eps C_{f'}(f)$.
\end{enumerate}
\end{lemma}
\begin{proof}
For each population $i \in K$, we have $\sum_{S\in\mathcal{S}_i}f_{i,S} = \sum_{S\in\mathcal{S}_i}f'_{i,S} = r_i$.
For the first statement, assume that $f$ is a Nash equilibrium.
Then every strategy $S \in \mathcal{S}_i$ with $f_{i,S} > 0$ has minimum latency.
Therefore,
\[
\sum_{S\in\mathcal{S}_i}f_{i,S}\ell_S(f) = r_i\min_{S\in\mathcal{S}_i}\ell_S(f) \le \sum_{S\in\mathcal{S}_i}f'_{i,S}\ell_S(f).
\]
Then summing over all populations gives
\[
C(f) = \sum_{i\in K}\sum_{S\in \mathcal{S}_i}f_{i,S}\ell_S(f) \le \sum_{i\in K}\sum_{S\in \mathcal{S}_i}f'_{i,S}\ell_S(f) = C_f(f').
\]

For the second statement, assume that $f'$ is an $\eps$-NE.
Then every strategy $S \in \mathcal{S}_i$ with $f'_{i,S} > 0$ has latency at most $1 + \eps$ times the minimum latency for population $i$.
Hence
\[
\sum_{S\in\mathcal{S}_i}f'_{i,S}\ell_S(f') \le (1 + \eps)r_i\min_{S\in\mathcal{S}_i}\ell_S(f') \le (1 + \eps)\sum_{S\in\mathcal{S}_i}f_{i,S}\ell_S(f').
\]
Then summing over all populations gives
\[
C(f') = \sum_{i\in K}\sum_{S\in \mathcal{S}_i}f'_{i,S}\ell_S(f') \le \sum_{i\in K}(1 + \eps)\sum_{S\in \mathcal{S}_i}f_{i,S}\ell_S(f') = (1 + \eps)C_{f'}(f).
\]
Finally, when both assumptions hold, subtracting the first inequality from the second gives
\[
\sum_{e\in E} (x'_e - x_e)\cdot (\ell_e(x'_e) - \ell_e(x_e)) = (C(f') - C_{f'}(f)) - (C_f(f') - C(f)) \le \eps C_{f'}(f). \qedhere
\]
\end{proof}

\subsection{Series-Parallel Networks}

Some of our results concern nonatomic routing games on series-parallel networks, by which we always mean games with a single commodity routed from the source to the sink of the network.
A series-parallel network $G$ with source $s_G$ and sink $t_G$ is defined recursively using two composition operations.
Formally, $G$ is series-parallel if it can be obtained from the following rules:
\begin{itemize}
    \item \textbf{Base case:} $G$ consists of a single edge from $s_G$ to $t_G$.
    \item \textbf{Series composition:} $G$ is obtained from two series-parallel networks $H_1$ and $H_2$ by identifying $t_{H_1}$ with $s_{H_2}$.
    In this case, $s_G = s_{H_1}$ and $t_G = t_{H_2}$.
    \item \textbf{Parallel composition:} $G$ is obtained from two series-parallel networks $H_1$ and $H_2$ by identifying $s_{H_1}$ with $s_{H_2}$ and $t_{H_1}$ with $t_{H_2}$.
    In this case, $s_G = s_{H_1} = s_{H_2}$ and $t_G = t_{H_1} = t_{H_2}$.
\end{itemize}

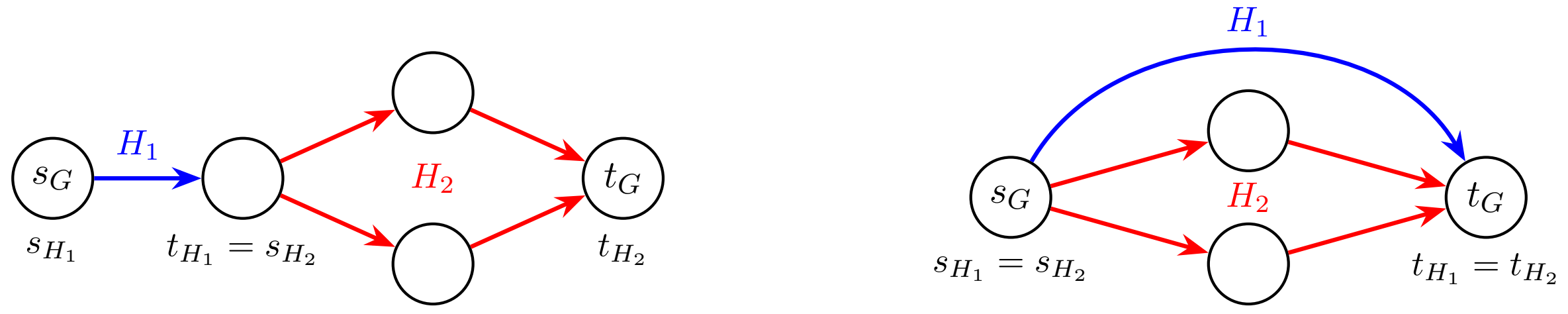
\begin{figure}[ht]
\centering
\begin{minipage}[b]{0.45\textwidth}
\centering
\begin{tikzpicture}[scale=0.95, auto, thick, >=Stealth]
    \node[node] (s) at (0,0) {$s_G$};
    \node[node] (m) at (2,0) {};
    \node[node] (a) at (4,0.9) {};
    \node[node] (b) at (4,-0.9) {};
    \node[node] (t) at (6,0) {$t_G$};

    \draw[->,blue,very thick] (s) -- (m);

    \draw[->,red,very thick] (m) -- (a);
    \draw[->,red,very thick] (a) -- (t);
    \draw[->,red,very thick] (m) -- (b);
    \draw[->,red,very thick] (b) -- (t);

    \node[font=\small] at (0,-0.75) {$s_{H_1}$};
    \node[font=\small] at (2,-0.75) {$t_{H_1} = s_{H_2}$};
    \node[font=\small] at (6,-0.75) {$t_{H_2}$};

    \node[font=\small,blue] at (0.9,0.35) {$H_1$};
    \node[font=\small,red] at (4,0) {$H_2$};
\end{tikzpicture}
\subcaption{Series composition}
\end{minipage}
\hfill
\begin{minipage}[b]{0.45\textwidth}
\centering
\begin{tikzpicture}[scale=0.95, auto, thick, >=Stealth]
    \node[node] (s) at (0,0) {$s_G$};
    \node[node] (a) at (2.5,0.7) {};
    \node[node] (b) at (2.5,-0.7) {};
    \node[node] (t) at (5,0) {$t_G$};

    \draw[->,blue,very thick]
        (s) to[bend left=60]
        node[midway,above,blue,font=\small] {$H_1$}
        (t);

    \draw[->,red,very thick] (s) -- (a);
    \draw[->,red,very thick] (a) -- (t);
    \draw[->,red,very thick] (s) -- (b);
    \draw[->,red,very thick] (b) -- (t);

    \node[font=\small,red] at (2.5,0) {$H_2$};

    \node[font=\small] at (0,-0.75) {$s_{H_1} = s_{H_2}$};
    \node[font=\small] at (5,-0.75) {$t_{H_1} = t_{H_2}$};
\end{tikzpicture}
\subcaption{Parallel composition}
\end{minipage}
\caption{Example of the two composition operations for series-parallel networks.}
\label{fig:sp_composition}
\end{figure}

Figure~\ref{fig:sp_composition} illustrates series and parallel composition.

For our results on series-parallel networks, we define the \emph{depth} $h_G$ of a network $G$ as the maximum number of edges on an $s_G$-$t_G$ path.
This notion of depth measures path length rather than the height of a series-parallel decomposition tree.
Equivalently, $h_G$ can be defined recursively.
If $G$ consists of a single edge, then $h_G = 1$.
If $G$ is a series composition of $H_1$ and $H_2$, then $h_G = h_{H_1} + h_{H_2}$.
If $G$ is a parallel composition of $H_1$ and $H_2$, then $h_G = \max\{h_{H_1}, h_{H_2}\}$.

\section{Tight Lower Bounds for \texorpdfstring{$\best$}{Best-ϵ-NE}}
\label{sec:best}

In this section, we give lower bounds on $\best$.
Our main technical result is a unified lower-bound theorem for polynomial latency functions (Theorem~\ref{thm:general_polynomial_meta_lowerbound}).
This theorem yields tight bounds for both general nonatomic congestion games and nonatomic routing games on series-parallel networks.
Its proof also recovers the classical price-of-anarchy bound.
We state the theorem and its consequences in Section~\ref{sec:unified_theorem} and prove it in Section~\ref{sec:unified_proof}.
We derive the bounds and give matching instances for general nonatomic congestion games in Section~\ref{sec:general_polynomial_lower} and for series-parallel networks in Section~\ref{sec:series_parallel_une_pne_polynomial_lower}.
Finally, in Section~\ref{sec:monotone_lower}, we consider arbitrary monotone latency functions.

\subsection{A Unified Theorem for Tight Bounds on \texorpdfstring{$\best$}{Best-ϵ-NE}}
\label{sec:unified_theorem}

We use the following two functions to state our bounds.

\begin{definition}
For every integer $d \ge 1$ and real number $x \in [0,1]$, define
\[
\mu_d(x) = x\left(1 - (1 - x)^{1/d}\right).
\]
As shown in Lemma~\ref{lem:mu_optima} in Section~\ref{sec:unified_proof}, $\mu_d$ is continuous, strictly increasing, and maps $[0,1]$ onto $[0,1]$.
Hence for every $\eps \ge 0$ we can define $\rho_d(\eps)$ to be the unique value in $[0,1)$ satisfying
\[
\mu_d(\rho_d(\eps)) = \eps(1 - \rho_d(\eps)).
\]
For example, when $d = 1$, we have $\mu_1(x) = x^2$ and $\rho_1(\eps) = \frac{\sqrt{\eps^2 + 4\eps} - \eps}{2}$.
\end{definition}

\begin{theorem}[Unified bound for polynomial latencies]
\label{thm:general_polynomial_meta_lowerbound}
Consider any nonatomic congestion game with polynomial latency functions of degree at most $d$.
Let $f$ be a Nash equilibrium and $f'$ an $\eps$-NE for some $\eps \ge 0$.
Suppose that $\frac{C(f) - C_{f'}(f)}{C(f)} \le \tau$ for some $\tau \ge 0$.
Then
\[
\frac{C(f')}{C(f)} \ge 1 - \theta + \mu_d(\theta),
\qquad
\theta = \min\left\{\tau, \rho_d(\eps), \frac{d}{d + 1}\right\}.
\]
\end{theorem}

For general congestion games, nonnegative latencies give $C_{f'}(f) \ge 0$, so we can take $\tau = 1$.

\begin{restatable}[General congestion games]{corollary}{corgeneralpolynomiallowerbound}
\label{cor:general_polynomial_lowerbound}
Consider any nonatomic congestion game with polynomial latency functions of degree at most $d$.
Let $f$ be a Nash equilibrium.
Then for every $\eps \ge 0$, every $\eps$-NE $f'$ satisfies
\[
\frac{C(f')}{C(f)} \ge 1 - \theta + \mu_d(\theta),
\qquad
\theta = \min\left\{\rho_d(\eps), \frac{d}{d + 1}\right\}.
\]
\end{restatable}

For series-parallel networks of depth $h$, Lemma~\ref{lem:series_parallel_usedpath_lowerbound} gives $C_{f'}(f) \ge \frac{C(f)}{1 + h\eps}$, so we can take $\tau = \frac{h\eps}{1 + h\eps}$.

\begin{restatable}[Series-parallel networks]{corollary}{corseriesparallelunepolynomiallowerbound}
\label{cor:series_parallel_une_polynomial_lowerbound}
Consider any nonatomic routing game on a depth-$h$ series-parallel network with polynomial latency functions of degree at most $d$.
Let $f$ be a Nash flow.
Then for every $\eps \ge 0$, every $\eps$-NE $f'$ satisfies
\[
\frac{C(f')}{C(f)} \ge 1 - \theta + \mu_d(\theta),
\qquad
\theta = \min\left\{\frac{h\eps}{1 + h\eps}, \rho_d(\eps), \frac{d}{d + 1}\right\}.
\]
\end{restatable}

As $\eps \to 0$ with $d$ and $h$ fixed, the general lower bound is $1 - (d\eps)^{1/2} + O_d(\eps)$, whereas the series-parallel lower bound is $1 - h\eps + O_{d,h}(\eps^2)$.
We give matching instances for the general and series-parallel bounds in Sections~\ref{sec:general_polynomial_lower} and~\ref{sec:series_parallel_une_pne_polynomial_lower}, respectively.

\begin{remark}[Recovering the price of anarchy]
\label{remark:poa}
The proof of Theorem~\ref{thm:general_polynomial_meta_lowerbound} yields the following bound for every flow $f'$, without assuming that it is an $\eps$-NE:
\[
\frac{C(f')}{C(f)} \ge \min_{x \in [0,1]}\bigl(1 - x + \mu_d(x)\bigr) = 1 - d(d + 1)^{-(d + 1)/d},
\]
where the equality follows from Lemma~\ref{lem:mu_optima}.
Taking $f'$ to be a socially optimal flow recovers the classical price-of-anarchy bound $\frac{1}{1 - d(d + 1)^{-(d + 1)/d}}$~\cite{Roughgarden02_poa}.
\end{remark}

\subsection{Proof of Theorem~\ref{thm:general_polynomial_meta_lowerbound}}
\label{sec:unified_proof}

The proof of Theorem~\ref{thm:general_polynomial_meta_lowerbound} relies on the following analytic properties of $\mu_d$, whose proofs are deferred to Appendix~\ref{sec:best_deferred}.

\begin{restatable}{lemma}{lemmuoptima}
\label{lem:mu_optima}
For every integer $d \ge 1$, $\mu_d$ has the following two properties:
\begin{enumerate}
    \item $\mu_d$ is continuous, strictly increasing and convex on $[0,1]$, and $\mu_d([0,1]) = [0,1]$.
    \item $1 - x + \mu_d(x)$ is decreasing on $[0,d/(d + 1)]$ and increasing on $[d/(d + 1),1]$, so $d/(d + 1)$ is the unique minimizer on $[0,1]$.
\end{enumerate}
\end{restatable}

\begin{proofof}{Theorem~\ref{thm:general_polynomial_meta_lowerbound}}
We compare $C(f)$ with $C(f')$ through the intermediate cost $C_{f'}(f)$ of the flow $f$ under the latencies induced by $f'$:
\[
C(f) - C(f') = \underbrace{\big(C(f) - C_{f'}(f)\big)}_{T} - \underbrace{\big(C(f') - C_{f'}(f)\big)}_{R}.
\]
Changing the latencies while keeping $f$ fixed reduces the cost by $T$, and then changing the flow to $f'$ increases the cost by $R$.
Let $x$ and $x'$ be the resource-load vectors induced by $f$ and $f'$, respectively.
\begin{align*}
T &= \sum_{e\in E}x_e(\ell_e(x_e) - \ell_e(x'_e)),\\
R &= \sum_{e\in E}(x'_e - x_e)\ell_e(x'_e).
\end{align*}

Therefore, lower bounding $C(f')$ reduces to upper bounding $T - R$, which we do in two steps.
We first use the equilibrium condition for $f$ and degree-$d$ polynomial latencies to lower bound $R$ in terms of $T$.
We then use the approximate-equilibrium condition for $f'$ to upper bound $T$.

To relate $T$ and $R$, define
\[
D = \sum_{e\in E}(x'_e - x_e)(\ell_e(x'_e) - \ell_e(x_e)).
\]
First, we have
\begin{align*}
R = \sum_{e\in E}(x'_e - x_e)\ell_e(x'_e)
&= \sum_{e\in E}(x'_e - x_e)\ell_e(x_e) + \sum_{e\in E}(x'_e - x_e)(\ell_e(x'_e) - \ell_e(x_e))\\
&= C_f(f') - C(f) + D \ge D \ge 0.
\end{align*}
The last two inequalities follow from Lemma~\ref{lem:residual_inequality} and the monotonicity of the latency functions.

If $T \le 0$, then $C(f') = C(f) - T + R \ge C(f)$, so the theorem holds.
For the rest of the proof, we assume $T > 0$.

Let $E_+$ be the set of resources $e$ with $x_e > x'_e$ and $\ell_e(x_e) > 0$.
Since every term of $T$ outside $E_+$ is nonpositive, we have
\[
T \le \sum_{e\in E_+}x_e(\ell_e(x_e) - \ell_e(x'_e)) = \sum_{e\in E_+}x_e\ell_e(x_e)\cdot \left(1 - \frac{\ell_e(x'_e)}{\ell_e(x_e)}\right).
\]
For $e \in E_+$, the contribution to $D$ is the contribution to $T$ multiplied by $1 - \frac{x'_e}{x_e}$:
\[
(x_e - x'_e)(\ell_e(x_e) - \ell_e(x'_e)) = x_e\ell_e(x_e)\cdot \left(1 - \frac{x'_e}{x_e}\right)\cdot \left(1 - \frac{\ell_e(x'_e)}{\ell_e(x_e)}\right).
\]
Suppose the latency function of resource $e$ is $\ell_e(x) = \sum_{j=0}^d a_{e,j}x^j$, where $a_{e,j} \ge 0$ for $0 \le j \le d$.
Then for each $e \in E_+$,
\[
\ell_e(x'_e) = \sum_{j=0}^d a_{e,j} (x'_e)^j = \sum_{j=0}^d a_{e,j} x_e^j \cdot \left(\frac{x'_e}{x_e}\right)^j
\ge \left(\frac{x'_e}{x_e}\right)^d\sum_{j=0}^d a_{e,j}x_e^j = \left(\frac{x'_e}{x_e}\right)^d\ell_e(x_e).
\]
Let $q_e = 1 - \frac{\ell_e(x'_e)}{\ell_e(x_e)}$ denote the relative decrease in latency on resource $e$.
Then
\begin{equation}
\label{eqn:unified_proof_tmp}
1 - q_e = \frac{\ell_e(x'_e)}{\ell_e(x_e)} \ge \left(\frac{x'_e}{x_e}\right)^d \implies 1 - \frac{x'_e}{x_e} \ge 1 - (1 - q_e)^{1/d}.
\end{equation}

By Lemma~\ref{lem:mu_optima}, $\mu_d$ is increasing and convex.
The weights $\frac{x_e\ell_e(x_e)}{C(f)}$ for $e \in E_+$ sum to at most $1$, so monotonicity and Jensen's inequality give
\begin{align*}
\mu_d\left(\frac{T}{C(f)}\right) &\le \mu_d\left(\sum_{e\in E_+}\frac{x_e\ell_e(x_e)}{C(f)}\cdot q_e + \left(1 - \sum_{e\in E_+}\frac{x_e\ell_e(x_e)}{C(f)}\right)\cdot 0\right)\\
&\le \sum_{e\in E_+}\frac{x_e\ell_e(x_e)}{C(f)}\cdot \mu_d(q_e) + \left(1 - \sum_{e\in E_+}\frac{x_e\ell_e(x_e)}{C(f)}\right)\mu_d(0)\\
&= \sum_{e\in E_+}\frac{x_e\ell_e(x_e)}{C(f)}\cdot q_e(1 - (1 - q_e)^{1/d}).
\end{align*}
Then plugging in $q_e = 1 - \frac{\ell_e(x'_e)}{\ell_e(x_e)}$ and inequality~\eqref{eqn:unified_proof_tmp} gives
\begin{align*}
\mu_d\left(\frac{T}{C(f)}\right) &\le \sum_{e\in E_+}\frac{x_e\ell_e(x_e)}{C(f)}\left(1 - \frac{\ell_e(x'_e)}{\ell_e(x_e)}\right)\left(1 - \frac{x'_e}{x_e}\right)\\
&= \frac{1}{C(f)}\sum_{e\in E_+}(x_e - x'_e)(\ell_e(x_e) - \ell_e(x'_e))\\
&\le \frac{D}{C(f)} \le \frac{R}{C(f)}.
\end{align*}
This allows us to establish a lower bound on $\frac{C(f')}{C(f)}$ that depends only on $T$:
\[
\frac{C(f')}{C(f)} = 1 - \frac{T}{C(f)} + \frac{R}{C(f)} \ge 1 - \frac{T}{C(f)} + \mu_d\left(\frac{T}{C(f)}\right).
\]

We now use the approximate-equilibrium condition for $f'$.
Lemma~\ref{lem:residual_inequality} gives $D \le \eps C_{f'}(f)$, so
\[
\mu_d\left(\frac{T}{C(f)}\right) \le \frac{D}{C(f)} \le \eps\frac{C_{f'}(f)}{C(f)} = \eps\left(1 - \frac{T}{C(f)}\right).
\]
By the definition and monotonicity of $\rho_d$, this implies $\frac{T}{C(f)} \le \rho_d(\eps)$.

By the second part of Lemma~\ref{lem:mu_optima}, the function $1 - x + \mu_d(x)$ is decreasing on $[0, \frac{d}{d + 1}]$ and is minimized at $x = \frac{d}{d + 1}$.
Since $\frac{T}{C(f)} \le \tau$ by assumption and $\frac{T}{C(f)} \le \rho_d(\eps)$, we conclude that
\[
\frac{C(f')}{C(f)} \ge 1 - \frac{T}{C(f)} + \mu_d\left(\frac{T}{C(f)}\right) \ge 1 - \theta + \mu_d(\theta),\qquad \theta = \min\left\{\tau, \rho_d(\eps), \frac{d}{d+1}\right\}. \qedhere
\]
\end{proofof}

\subsection{General Nonatomic Congestion Games with Polynomial Latencies}
\label{sec:general_polynomial_lower}

\corgeneralpolynomiallowerbound*
\begin{proof}
Nonnegative latencies imply $\frac{C(f) - C_{f'}(f)}{C(f)} \le 1$.
The corollary follows from Theorem~\ref{thm:general_polynomial_meta_lowerbound} with $\tau = 1$, since $\frac{d}{d + 1} < 1$.
\end{proof}

As $x \to 0$, we have $\mu_d(x) = x^2/d + O_d(x^3)$, and hence $\rho_d(\eps) = \sqrt{d\eps} + O_d(\eps)$ as $\eps \to 0$.
Therefore,
\[
\frac{C(f')}{C(f)} \ge 1 - \rho_d(\eps) + \eps(1 - \rho_d(\eps)) = 1 - \sqrt{d\eps} + O_d(\eps).
\]

We next show that Corollary~\ref{cor:general_polynomial_lowerbound} is tight for every $\eps > 0$.
Let $\eps_0$ be the unique value satisfying $\rho_d(\eps_0) = \frac{d}{d + 1}$.
It suffices to construct matching instances for $\eps \le \eps_0$, where $\theta = \rho_d(\eps)$.
For $\eps > \eps_0$, we have $\theta = \frac{d}{d + 1}$, so the bound in Corollary~\ref{cor:general_polynomial_lowerbound} equals its value at $\eps_0$, and the matching instances for $\eps_0$ remain tight because every $\eps_0$-NE is an $\eps$-NE.
We use a family of instances with the same combinatorial structure as the tight instance for approximate PoA in Christodoulou et~al.~\cite{ChristodoulouKS11}, with different latency functions.

\begin{example}
\label{ex:general_polynomial_lowerbound_tight}
Fix $\eps > 0$ such that $\theta = \rho_d(\eps) \le \frac{d}{d+1}$.
Take integers $m \ge k \ge 1$ such that $k/m$ gives an arbitrarily good approximation to $(1 - \theta)^{1/d}$.
Below we simply assume that $k/m = (1 - \theta)^{1/d}$.

The instance consists of $2(m + k)$ resources $a_0, \ldots, a_{m+k-1}, b_0, \ldots, b_{m+k-1}$.
Each resource $a_j$ has latency $\ell_{a_j}(x) = x^d$, and each resource $b_j$ has constant latency $\ell_{b_j}(x) = \gamma = (m - k)m^d$.

There are $m + k$ populations, each with demand one.
Population $i$ has the two strategies
\[
A_i = \{a_i,a_{i+1},\ldots,a_{i+m-1}\},\qquad
P_i = \{a_{i+m},a_{i+m+1},\ldots,a_{i+m+k-1},b_i\}.
\]
The indices are taken modulo $m + k$.
In words, $a_0, \ldots, a_{m+k-1}$ lie on a circle and each $A_i$ corresponds to a unique cyclic interval of $m$ resources, while $P_i$ consists of the remaining $k$ resources and a private resource $b_i$.

Let $f$ be a flow that assigns the entire demand of every population $i$ to $A_i$.
Then every $a_j$ has load $m$.
Consequently, the used strategies $A_i$ each have cost $m\cdot m^d = m^{d+1}$, while the alternative $P_i$ has cost $km^d + \gamma = m^{d+1}$, so $f$ is indeed a Nash equilibrium.

Next consider a flow $f'$ that assigns the entire demand of every population $i$ to $P_i$.
Then every $a_j$ has load $k$, so each used strategy $P_i$ has cost $k^{d+1} + \gamma$, while the alternative $A_i$ has cost $mk^d$.
Using $k/m = (1 - \theta)^{1/d} \iff (k/m)^d = 1 - \theta$, we have
\[
\frac{k^{d+1}+\gamma}{mk^d} = \frac{(k/m)^{d+1} + 1-(k/m)}{(k/m)^d} = 1 + \frac{(1-k/m)(1-(k/m)^d)}{(k/m)^d} = 1 + \frac{\mu_d(\theta)}{1-\theta} = 1 + \eps.
\]
Hence $f'$ is an $\eps$-NE.
Finally, since all populations have the same cost within each flow, we have
\[
\frac{C(f')}{C(f)}
= \frac{k^{d+1}+\gamma}{m^{d+1}}
= \left(\frac{k}{m}\right)^{d+1} + 1 - \frac{k}{m}
= 1 - \theta + \theta\left(1 - (1 - \theta)^{1/d}\right)
= 1 - \theta + \mu_d(\theta),
\]
which exactly matches the bound in Corollary~\ref{cor:general_polynomial_lowerbound}.
\end{example}

Corollary~\ref{cor:general_polynomial_lowerbound} is tight even for single-commodity routing games: a family of Braess graphs achieves the bound exactly for a sequence of values of $\eps$ tending to $0$.
The instance and its analysis are presented in Appendix~\ref{sec:braess_tight_examples}.

\subsection{Series-Parallel Networks with Polynomial Latencies}
\label{sec:series_parallel_une_pne_polynomial_lower}

In Section~\ref{sec:general_polynomial_lower}, we showed that for degree-$d$ polynomial latencies, the largest possible relative reduction in social cost is of order $\sqrt{d\eps}$, even for single-commodity routing games on Braess graphs.
In this subsection, we show that for series-parallel networks of depth $h$, this reduction is only of order $h\eps$.
This contrasts with the classical PoA bound, which is already tight on a network of two parallel edges and therefore remains unchanged when restricted to series-parallel networks.

The next lemma captures a key structural property of $\eps$-NE in series-parallel networks.
Its proof uses the observation that if an $\eps$-NE $f'$ routes at least as much total traffic as a Nash flow $f$ through two subnetworks in parallel, then it does so through at least one of them.
The proof proceeds by induction on the recursive structure of the network and is deferred to Appendix~\ref{sec:series_parallel_deferred}.

\begin{restatable}{lemma}{lemseriesparallelusedpathlowerbound}
\label{lem:series_parallel_usedpath_lowerbound}
Let $G$ be a series-parallel network of depth $h_G$ with monotone latency functions.
Let $f$ be a Nash flow, and let $L_G$ be the common latency of the paths used by $f$.
Let $f'$ be an $\eps$-NE.
Then every $s_G$-$t_G$ path under $f'$ has latency at least $\frac{L_G}{1 + h_G\eps}$.
\end{restatable}

\corseriesparallelunepolynomiallowerbound*
\begin{proof}
Let $L_G$ be the latency of the paths used by $f$.
By Lemma~\ref{lem:series_parallel_usedpath_lowerbound}, every such path has latency at least $\frac{L_G}{1 + h\eps}$ under $f'$, so $C_{f'}(f) \ge \frac{C(f)}{1 + h\eps}$.
The corollary follows from Theorem~\ref{thm:general_polynomial_meta_lowerbound} with $\tau = \frac{h\eps}{1 + h\eps}$.
\end{proof}

For fixed $d$ and $h$, we have $\theta = \frac{h\eps}{1 + h\eps}$ for sufficiently small $\eps$, so Corollary~\ref{cor:series_parallel_une_polynomial_lowerbound} gives $\frac{C(f')}{C(f)} \ge 1 - h\eps + O_{d,h}(\eps^2)$ as $\eps \to 0$.
Thus, to first order, the relative reduction in social cost is at most $h\eps$.

We next show that Corollary~\ref{cor:series_parallel_une_polynomial_lowerbound} is tight for every $\eps > 0$, with the depth $h$ chosen as part of the construction in Case 2.
Let $\eps_0$ be the unique value satisfying $\rho_d(\eps_0) = \frac{d}{d + 1}$.
Case 2 below matches the bound at $\eps_0$, where $\theta = \frac{d}{d + 1}$.
These instances remain tight for every $\eps > \eps_0$, since the bound remains constant on these networks and every $\eps_0$-NE is an $\eps$-NE.

\begin{example}
\label{ex:h-pigou_polynomial}
We give two families of instances with unit demand, corresponding to the first two cases of $\theta$ in Corollary~\ref{cor:series_parallel_une_polynomial_lowerbound}.

\paragraph{Case 1: $\theta = \frac{h\eps}{1+h\eps}$.}
Thus $\theta \le \rho_d(\eps)$ and $\theta \le \frac{d}{d+1}$.
The instance is a concatenation of $h$ copies of the following Pigouvian network: the upper edge has latency function $\ell(x) = x^d/h$ and the lower edge has constant latency $1/h$.
Figure~\ref{fig:k-pigou_polynomial} provides an illustration of the instance for $h = 4$.

\begin{figure}[ht]
\centering
\begin{tikzpicture}[scale=1.0, auto, thick, >=Stealth]
    \node[node] (s) at (0,0) {$s$};
    \node[node] (v1) at (3,0) {};
    \node[node] (v2) at (6,0) {};
    \node[node] (v3) at (9,0) {};
    \node[node] (t) at (12,0) {$t$};

    \draw[->,edge] (s) to[bend left=28] node[midway, above] {$x^d/4$} (v1);
    \draw[->,edge] (s) to[bend right=28] node[midway, below] {$1/4$} (v1);

    \draw[->,edge] (v1) to[bend left=28] node[midway, above] {$x^d/4$} (v2);
    \draw[->,edge] (v1) to[bend right=28] node[midway, below] {$1/4$} (v2);

    \draw[->,edge] (v2) to[bend left=28] node[midway, above] {$x^d/4$} (v3);
    \draw[->,edge] (v2) to[bend right=28] node[midway, below] {$1/4$} (v3);

    \draw[->,edge] (v3) to[bend left=28] node[midway, above] {$x^d/4$} (t);
    \draw[->,edge] (v3) to[bend right=28] node[midway, below] {$1/4$} (t);

    \draw[->,very thick,blue] (s) to[bend left=28] (v1);
    \draw[->,very thick,blue] (v2) to[bend left=28] (v3);
    \draw[->,very thick,blue] (v3) to[bend left=28] (t);
    \draw[->,very thick,blue] (v1) to[bend right=28] (v2);
\end{tikzpicture}
\caption{An illustration of the tight instance of Case 1, where $h = 4$. The blue path is one of the used paths in the $\eps$-NE.}
\label{fig:k-pigou_polynomial}
\end{figure}
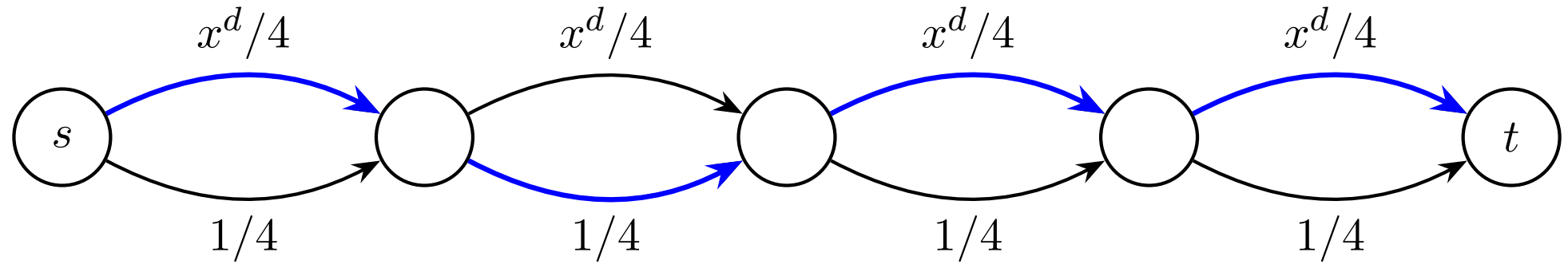

The Nash flow sends all traffic through the upper edges and has cost $1$.

Set $\delta = \mu_d(\theta) / \theta = 1 - (1 - \theta)^{1/d}$.
We have $\theta \le \rho_d(\eps) \implies \mu_d(\theta) \le \eps(1 - \theta) = \theta/h$, which implies $\delta \le 1/h$.
Consider the flow $f'$ which routes $1 - h\delta$ units of traffic on the all-upper path, and for each $1 \le i \le h$, routes $\delta$ units of traffic on the path that uses the lower edge in the $i$-th copy and the upper edge in all other copies.
Every upper edge has load $1 - \delta$, thus the all-upper path has latency $(1 - \delta)^d = 1 - \theta$ and is the shortest path, while all other $h$ used paths have latency $1 - \theta + \theta/h = (1 + \eps)(1 - \theta)$, hence $f'$ is indeed an $\eps$-NE.
Its social cost is
\[
(1 - h\delta)(1 - \theta) + h\delta(1 - \theta + \theta/h) = 1 - \theta + \delta\theta = 1 - \theta + \mu_d(\theta).
\]

\paragraph{Case 2: $\theta = \rho_d(\eps)$.}
Fix $\eps > 0$ such that $\theta = \rho_d(\eps) \le d/(d + 1)$.
Take integers $h \ge 2$ and $1 \le k < h$ such that $k/h$ gives an arbitrarily good approximation to $1 - (1 - \theta)^{1/d}$.
Below we simply assume that $k/h = 1 - (1 - \theta)^{1/d} = \mu_d(\theta)/\theta$.
Since $\theta = \rho_d(\eps)$, we have $\mu_d(\theta) = \eps(1 - \theta)$ and hence $h\eps = k\theta/(1 - \theta)$.
In particular, $k \ge 1$ implies $\theta \le h\eps/(1 + h\eps)$, so $\theta$ is indeed the active term in the corollary.

Use the same $h$-layer network as in Case 1.
The Nash flow sends all traffic through the upper edges and has cost $1$.
For each $1 \le j \le h$, let $P_j$ be the path that uses the lower edge in layers $j, j + 1, \ldots, j + k - 1$, with indices taken modulo $h$, and the upper edge in every other layer.
Consider the flow $f'$ that routes $1/h$ units of traffic on each $P_j$.
Every upper edge is used by exactly $h - k$ of these paths and therefore has load $1 - k/h = (1 - \theta)^{1/d}$ and latency $(1 - \theta)/h$.
Since the lower edge in each layer has latency $1/h$, the all-upper path is a shortest path and has latency $1 - \theta$, while every used path $P_j$ has latency
\[
\frac{h-k}{h}(1 - \theta) + \frac{k}{h}
= 1 - \theta + \frac{k}{h}\theta
= 1 - \theta + \mu_d(\theta)
= (1 + \eps)(1 - \theta).
\]
Hence $f'$ is an $\eps$-NE.
Since every used path has the same latency and the total demand is one, its social cost is $C(f') = 1 - \theta + \mu_d(\theta)$.
\end{example}

\subsection{Arbitrary Monotone Latencies}
\label{sec:monotone_lower}

We now consider arbitrary monotone latency functions.
For general nonatomic congestion games and any fixed $\eps > 0$, there are instances with a Nash equilibrium $f$ and an $\eps$-NE $f'$ for which $\frac{C(f')}{C(f)}$ is arbitrarily close to $0$.
This holds even for single-commodity routing games, as we show in Example~\ref{ex:braess_polynomial_lowerbound_tight} in Appendix~\ref{sec:braess_tight_examples}.

In contrast, for series-parallel networks of depth $h$, the following theorem gives a tight lower bound of $\frac{1}{1 + h\eps}$ on $\best$.

\begin{theorem}
\label{thm:series_parallel_une_monotone_lowerbound}
Consider any nonatomic routing game on a depth-$h$ series-parallel network with monotone latency functions.
Let $f$ be a Nash flow.
Then for every $\eps \ge 0$, every $\eps$-NE $f'$ satisfies
\[
\frac{C(f')}{C(f)} \ge \frac{1}{1 + h\eps}.
\]
\end{theorem}
\begin{proof}
Let $r$ be the demand and let $L_G$ be the common latency of the paths used by $f$, so $C(f) = rL_G$.
By Lemma~\ref{lem:series_parallel_usedpath_lowerbound}, every path under $f'$ has latency at least $\frac{L_G}{1 + h\eps}$, so $C(f') \ge \frac{rL_G}{1 + h\eps} = \frac{C(f)}{1 + h\eps}$.
\end{proof}

The following variant of Case 1 in Example~\ref{ex:h-pigou_polynomial} shows the tightness of Theorem~\ref{thm:series_parallel_une_monotone_lowerbound}.

\begin{example}
\label{ex:h-pigou_monotone}
Fix an integer $h \ge 1$ and $\delta \in (0,1/h)$.
Consider a series composition of $h$ copies of a network consisting of two parallel edges:
The upper edge has latency $1/(h(1 + h\eps))$ at loads at most $1 - \delta$, latency $1/h$ at loads at least $1$, and is linear in between.
The lower edge has constant latency $1/h$.
For unit demand, the Nash flow sends all traffic on the all-upper path and has cost $1$.

Consider the flow $f'$ that routes $1 - h\delta$ units of traffic on the all-upper path and, for each $1 \le i \le h$, routes $\delta$ units of traffic on the path that uses the lower edge in the $i$-th copy and the upper edge in all other copies.
Under $f'$, every upper edge has load $1 - \delta$ and latency $1/(h(1 + h\eps))$.
Therefore, the all-upper path has latency $1/(1 + h\eps)$ and is a shortest path, while each of the remaining $h$ used paths has latency $(1 + \eps)/(1 + h\eps)$.
Thus $f'$ is an $\eps$-NE with social cost
\[
C(f') = (1 - h\delta)\frac{1}{1 + h\eps}
+ h\delta\frac{1 + \eps}{1 + h\eps}
= \frac{1 + h\delta\eps}{1 + h\eps},
\]
which tends to $\frac{1}{1 + h\eps}$ as $\delta \to 0$.
\end{example}

\section{Tight Upper Bounds for \texorpdfstring{$\worst$}{Worst-ϵ-NE}}
\label{sec:worst}

In this section we provide tight upper bounds on $\worst$.

\subsection{Tight Bounds on \texorpdfstring{$\worst$}{Worst-ϵ-NE} via Lagrangian Duality}

When $\eps$ is small, the upper bound is governed by the following functions:

\begin{definition}
\label{def:lambda}
For every integer $d \ge 1$ and real number $x \in [0,+\infty)$, define
\[
\lambda_d(x) = \frac{x}{1+x}\left(1 - (1 + x)^{-1/d}\right).
\]
In Lemma~\ref{lem:lambda_monotonicity} we justify that $\lambda_d$ is continuous, strictly increasing, and maps $[0,+\infty)$ onto $[0,1)$, so we can define its inverse function $\omega_d$: for every $x \in [0,1)$, $\omega_d(x) = \lambda_d^{-1}(x)$.

For example, when $d = 1$, we have $\lambda_1(x) = \frac{x^2}{(1+x)^2}$ and $\omega_1(x) = \frac{\sqrt{x}}{1-\sqrt{x}}$.
\end{definition}

Define the constant $c_d$ and the thresholds $\eps_1, \eps_2$ as follows:
\[
c_d := \frac{d}{(d+1)^{(d+1)/d}},\qquad
\eps_1 := \frac{\lambda_d(d)}{1-\lambda_d(d)}
= \frac{d\left((d+1)^{1/d}-1\right)}{d+(d+1)^{1/d}},\qquad
\eps_2 := (d + 1)^{1/d} - 1.
\]

\begin{theorem}
\label{thm:general_polynomial_upperbound}
Consider any nonatomic congestion game with polynomial latency functions of degree at most $d$.
Let $f$ be a Nash equilibrium.
Then for every $\eps \ge 0$, every $\eps$-NE $f'$ satisfies
\[
\frac{C(f')}{C(f)} \le
\begin{cases}
\displaystyle 1 + \omega_d\left(\frac{\eps}{1+\eps}\right), & 0 \le \eps \le \eps_1, \\[2ex]
\displaystyle \left(\frac{1}{1+\eps} - c_d\right)^{-1}, & \eps_1 \le \eps \le \eps_2, \\[2ex]
(1 + \eps)^{d+1}, & \eps \ge \eps_2.
\end{cases}
\]
\end{theorem}

We deliberately include the splitting points $\eps_1, \eps_2$ twice to emphasize that the upper bound is continuous: the bounds for the first and second regimes coincide at $\eps_1$ and the bounds for the second and third regimes coincide at $\eps_2$.

As a concrete example, when $d = 1$, the upper bound is
\[
\frac{C(f')}{C(f)} \le
\begin{cases}
\displaystyle \frac{1}{1-\sqrt{\eps/(1+\eps)}}, & 0 \le \eps \le \frac{1}{3}, \\[2ex]
\displaystyle \frac{4(1+\eps)}{3-\eps}, & \frac{1}{3} \le \eps \le 1, \\[2ex]
(1 + \eps)^2, & \eps \ge 1.
\end{cases}
\]
For fixed $d$, as $\eps \to 0$, the first regime gives
\[
\frac{C(f')}{C(f)} \le 1 + \sqrt{d\eps} + O_d(\eps).
\]

\begin{proofof}{Theorem~\ref{thm:general_polynomial_upperbound}}
Let $R = C(f') - C_{f'}(f)$ and $\alpha = \frac{C(f')}{C(f)}$.
When $\alpha \le 1$ the claimed bounds are already satisfied, so we focus on $\alpha > 1$.
By Lemma~\ref{lem:residual_inequality}, we have
\[
\frac{R}{C(f)} = \frac{R}{C(f')}\alpha = \frac{C(f')-C_{f'}(f)}{C(f')}\alpha \le \frac{\eps}{1+\eps}\alpha.
\]
To upper bound $\alpha$, we formulate an optimization problem parameterized by $\alpha$, which produces lower bounds on $\frac{R}{C(f)}$ as a function of $\alpha$.
For example, for the first regime $0 \le \eps \le \eps_1$, we will eventually show
\[
\alpha\lambda_d(\alpha - 1) \le \frac{R}{C(f)} \le \frac{\eps}{1+\eps}\alpha,
\]
which by Definition~\ref{def:lambda} implies $\alpha \le 1 + \omega_d\left(\frac{\eps}{1+\eps}\right)$.

\paragraph{The Lagrangian dual.}
We formulate the following constrained optimization program, which minimizes $R = C(f') - C_{f'}(f)$ subject to the constraint that $f'$ has social cost $\alpha C(f)$ and an inequality implied by the equilibrium condition for $f$.
Let $x$ and $x'$ denote the resource-load vectors induced by $f$ and $f'$, respectively.
\begin{lp}
    \mini{\sum_{e\in E}(x'_e - x_e)\ell_e(x'_e)} \\
    \st \con{\sum_{e\in E}x'_e \ell_e(x'_e) = \alpha C(f)} \\
    \con{\sum_{e\in E} (x'_e - x_e)\ell_e(x_e) \ge 0} \\
    \con{\sum_{e\in E} x_e\ell_e(x_e) = C(f)}\\
    \qcon{x_e \ge 0, x'_e \ge 0}{\text{all $e \in E$}}
\tag{\text{P}}\label{lp:lowerbound}
\end{lp}%
Then any lower bound on the optimal value of~\eqref{lp:lowerbound} is a valid lower bound for $R$.

We derive a lower bound on the objective value of~\eqref{lp:lowerbound} via Lagrangian duality.
Let $A \ge 0$ be the Lagrangian multiplier for $\sum_e (x'_e - x_e)\ell_e(x_e) \ge 0$, and $\eta \in \mathbb{R}$ the multiplier for $\sum_{e\in E}x'_e\ell_e(x'_e) = \alpha C(f)$.
We keep $\sum_{e\in E}x_e\ell_e(x_e) = C(f)$ as a domain constraint that is not dualized.
The Lagrangian is
\[
\mathcal{L}(x,x';A,\eta) = \sum_{e\in E}(x'_e - x_e)\ell_e(x'_e) - A\left(\sum_e (x'_e - x_e)\ell_e(x_e)\right) - \eta\left(\sum_{e\in E}x'_e\ell_e(x'_e) - \alpha C(f)\right).
\]
Suppose the latency function of each resource $e \in E$ is a degree-$d$ polynomial $\ell_e(x) = \sum_{i=0}^d a_{e,i}x^i$, and set $B = 1 - \eta$.
Then
\[
\mathcal{L}(x,x';A,\eta) = \sum_{e\in E}\sum_{i=0}^d a_{e,i}\left[B(x'_e)^{i+1} - x_e(x'_e)^i - A(x'_e)x_e^i + Ax_e^{i+1}\right] + (1 - B)\alpha C(f).
\]
If there is a scalar $t \in \mathbb{R}$ such that
\[
B(x'_e)^{i+1} - x_e(x'_e)^i - A(x'_e)x_e^i + Ax_e^{i+1} + tx_e^{i+1} \ge 0
\]
holds for all $e \in E$ and $0 \le i \le d$, then we have
\begin{align*}
\sum_{e\in E}(x'_e - x_e)\ell_e(x'_e)
\ge \mathcal{L}(x,x';A,\eta)
&\ge (1 - B)\alpha C(f) - t\sum_{e\in E}\sum_{i=0}^d a_{e,i}x_e^{i+1}\\
&= (1 - B)\alpha C(f) - t\sum_{e\in E}x_e\ell_e(x_e)\\
&= \left((1 - B)\alpha - t\right)C(f).
\end{align*}
Thus a lower bound on $\frac{R}{C(f)}$ reduces to the following program, which is independent of the game structure and the specific coefficients in the latency functions.
\stepcounter{equation}
\begin{lp}
    \maxi{(1 - B)\alpha - t} \\
    \st \qcon{By^{i+1} - xy^i - Ayx^i + Ax^{i+1} + tx^{i+1} \ge 0}{\text{all $x \ge 0, y \ge 0,\,i = 0, \ldots, d$}} \\
    \con{A \ge 0}
\tag{\text{D}}\label{lp:dual}
\end{lp}%
Every feasible solution $(A,B,t)$ of~\eqref{lp:dual} yields a lower bound $\frac{R}{C(f)} \ge (1 - B)\alpha - t$.

\paragraph{The special case $A = 0$.}
When $A = 0$, the dual constraints simplify to $By^{i+1} - xy^i + tx^{i+1} \ge 0$.
Taking $y = 0, i = 0$ gives $t \ge 1$, while $x = 0$ requires $B \ge 0$.

For $x > 0$, let $q = y/x$.
The dual constraints then become
\[
g_i(q) := Bq^{i+1} - q^i + t \ge 0 \qquad \text{for all }q \ge 0,\,i = 0, \ldots, d.
\]
Notice that for all $i = 0, \ldots, d - 1$,
\[
g_{i+1}(q) - g_i(q) = Bq^{i+2} - q^{i+1} - Bq^{i+1} + q^i = q^i(q - 1)(Bq - 1).
\]
For each fixed $q\ge0$, these differences are either all nonnegative or all nonpositive.
Thus for every fixed $q$, the minimum of $g_i(q)$ over $i = 0, \ldots, d$ is attained at either $i = 0$ or $i = d$.

Therefore, in order for $(A = 0,B,t)$ to be a feasible dual solution, it suffices to ensure that
\begin{enumerate}
    \item $g_0(q) = Bq - 1 + t \ge 0$ for all $q \ge 0$, which holds as long as $B \ge 0$.
    \item $g_d(q) = Bq^{d+1} - q^d + t \ge 0$ for all $q$, which holds as long as
        \[
        B \ge \sup_{q>0}\left\{\frac{1}{q} - \frac{t}{q^{d+1}}\right\}.
        \]
\end{enumerate}
For $t \ge 1$, the supremum is attained at $q = ((d + 1)t)^{1/d}$, which gives
\[
B \ge \frac{d}{(d+1)^{(d+1)/d}}t^{-1/d} = c_dt^{-1/d}.
\]
Since our goal is to maximize the dual objective $(1 - B)\alpha - t$, we focus on dual solutions where $B = c_dt^{-1/d}$.

The objective value of the dual solution $(A = 0,B = c_dt^{-1/d},t)$ is $\alpha - t - c_d\alpha t^{-1/d}$, and as argued above, it is a lower bound for $\frac{R}{C(f)}$.
Therefore,
\[
\alpha - t - c_d\alpha t^{-1/d} \le \frac{R}{C(f)} \le \frac{\eps}{1+\eps}\alpha,
\]
which gives $\alpha \le \left(\frac{1}{1+\eps} - c_dt^{-1/d}\right)^{-1}t =: U(t)$ provided that $\frac{1}{1+\eps} - c_dt^{-1/d} > 0$.
We minimize $U(t)$ over $t \ge 1$ satisfying this condition.
Differentiating gives
\[
U'(t) = \frac{\frac{1}{1+\eps}-\frac{d+1}{d}c_dt^{-1/d}}{\left(\frac{1}{1+\eps}-c_dt^{-1/d}\right)^2},
\]
whose numerator is strictly increasing on $t > 0$ and is equal to zero at $t^* = \frac{(1+\eps)^d}{d+1}$.
Thus, depending on whether $t^* \le 1$ or $t^* \ge 1$, we obtain different minimizers of $U(t)$ which correspond to the second and third regimes in the claimed bound.
\begin{itemize}
    \item When $\eps \le \eps_2$, $t^* \le 1$, so $\alpha \le U(1) = \left(\frac{1}{1+\eps} - c_d\right)^{-1}$.
    \item When $\eps \ge \eps_2$, $t^* \ge 1$, so $\alpha \le U(t^*) = (1 + \eps)^{d+1}$.
\end{itemize}

\paragraph{The first regime.}
We now allow $A \ge 0$ to vary and derive the upper bound for the first regime $0 \le \eps \le \eps_1$.
For $0 \le \eps \le \eps_1$, the preceding $A = 0$ bound gives $\alpha \le d + 1$.

When $x = 0$, the dual constraints become $By^{i+1} \ge 0$ for $i \ge 1$ and $(B - A)y \ge 0$ for $i = 0$, which are satisfied as long as $B \ge A \ge 0$.

For $x > 0$, let $q = y/x$.
The dual constraints become
\[
Bq^{i+1} - q^i - Aq + A + t \ge 0
\qquad\text{for all }q \ge 0,\,i = 0, \ldots, d.
\]
At $i = 0, q = 0$ this gives $A + t \ge 1$.
At $i = d, q > 0$ this gives
\begin{equation}
\label{eqn:upperbound_tmp1}
B \ge q^{-1} + Aq^{-d} - (A + t)q^{-d-1}.
\end{equation}
Substituting it into the dual objective yields an upper bound of
\begin{equation}
\label{eqn:upperbound_tmp2}
(1 - B)\alpha - t \le \alpha\left(1 - \frac{1}{q}\right)
- \frac{\alpha(q-1)}{q^{d+1}}A
+ \left(\frac{\alpha}{q^{d+1}} - 1\right)t.
\end{equation}
Evaluating this upper bound at $q = \alpha^{1/d}$, we obtain
\begin{equation}
\label{eqn:upperbound_tmp3}
(1 - B)\alpha - t \le (\alpha - A - t)(1 - \alpha^{-1/d})
\le (\alpha - 1)(1 - \alpha^{-1/d}) = \alpha\lambda_d(\alpha - 1).
\end{equation}

Next we construct a dual solution with objective value exactly $\alpha \lambda_d(\alpha - 1)$.
To achieve this, the above argument suggests that all inequalities in~\eqref{eqn:upperbound_tmp1},~\eqref{eqn:upperbound_tmp2}, and~\eqref{eqn:upperbound_tmp3} must be tight when $q = \alpha^{1/d}$.
This first implies $A + t = 1$, so from now on we fix $t = 1 - A$.
Then the dual constraints become
\[
h_i(q) := Bq^{i+1} - q^i - Aq + 1 \ge 0
\qquad\text{for all }q \ge 0,\,i = 0, \ldots, d.
\]
Notice that $h_{i+1}(q) - h_i(q) = g_{i+1}(q) - g_i(q)$.
By the same argument as in the case $A = 0$, for each fixed $q$ the minimum of $h_i(q)$ over $i = 0,\ldots,d$ is attained at either $i = 0$ or $i = d$.
Since $h_0(q) = (B - A)q \ge 0$ whenever $B \ge A$, it suffices to satisfy $h_d(q) \ge 0$ for all $q \ge 0$.

To make inequality~\eqref{eqn:upperbound_tmp1} tight at $q = \alpha^{1/d}$, we must have
\begin{equation}
\label{eqn:upperbound_tmp4}
h_d(\alpha^{1/d}) = 1 + B\alpha^{(d+1)/d} - A\alpha^{1/d} - \alpha = 0.
\end{equation}
Moreover, since we must ensure $h_d(q) \ge 0$ for all $q > 0$, $q = \alpha^{1/d}$ must be a local minimizer of $h_d$, so the first-order optimality condition requires
\begin{equation}
\label{eqn:upperbound_tmp5}
h'_d(\alpha^{1/d}) = (d + 1)B\alpha - A - d\alpha^{(d-1)/d} = 0.
\end{equation}
Solving equations~\eqref{eqn:upperbound_tmp4} and~\eqref{eqn:upperbound_tmp5} gives
\[
A = \frac{d+1-\alpha}{d\alpha^{1/d}},\qquad
B = \frac{(d-1)\alpha+1}{d\alpha^{(d+1)/d}}.
\]
The condition $A \ge 0$ restricts this construction to $\alpha \le d + 1$.

We verify that this is indeed a feasible dual solution.
We have shown that it suffices to check $B \ge A$ and $h_d(q) \ge 0$ for all $q \ge 0$.
We have
\[
B - A = \frac{(d-1)\alpha+1-\alpha(d+1-\alpha)}{d\alpha^{(d+1)/d}} = \frac{(\alpha-1)^2}{d\alpha^{(d+1)/d}} \ge 0.
\]
By construction, $h_d(\alpha^{1/d}) = h'_d(\alpha^{1/d}) = 0$.
Now $h'_d(q) = q^{d-1}((d + 1)Bq - d) - A$ is nonpositive when $q \le \frac{d}{(d+1)B}$.
Above that point, $h'_d(q)$ increases with $q$.
Since $A \ge 0$, the stationary point $\alpha^{1/d}$ lies in this latter interval, so the derivative is nonpositive before $\alpha^{1/d}$ and nonnegative after it.
Hence $h_d(q) \ge 0$ for all $q \ge 0$.

Finally, for $1 < \alpha \le d + 1$, by Lagrangian duality, this dual solution implies
\[
\alpha\lambda_d(\alpha - 1) \le \frac{R}{C(f)} \le \frac{\eps}{1+\eps}\alpha,
\]
which gives $\alpha \le 1 + \omega_d\left(\frac{\eps}{1+\eps}\right)$.
\end{proofof}

\subsection{Matching Instances}

We next show that the upper bound in the first regime is tight.
The instance has the same combinatorial structure as Example~\ref{ex:general_polynomial_lowerbound_tight}, with a different constant latency for the private resources.
We present the complete instance again for the reader's convenience.

\begin{example}
\label{ex:general_polynomial_upperbound_tight}
Fix $0 < \eps \le \eps_1$ and take integers $m \ge k \ge 1$ such that $m/k$ gives an arbitrarily good approximation to $\bigl(1 + \omega_d(\eps/(1 + \eps))\bigr)^{1/d}$.
Below we assume that they are equal.

The instance consists of $2(m + k)$ resources $a_0, \ldots, a_{m+k-1}, b_0, \ldots, b_{m+k-1}$.
Each resource $a_j$ has latency $x^d$ and each resource $b_j$ has constant latency $\gamma = (m - k)k^d$.
There are $m + k$ populations, each with demand one.
Population $i$ has the two strategies
\[
A_i = \{a_i,\ldots,a_{i+m-1}\},\qquad
P_i = \{a_{i+m},\ldots,a_{i+m+k-1},b_i\},
\]
with indices taken modulo $m + k$.

Let $f$ be a flow that assigns the entire demand of every population $i$ to $P_i$.
Then every $a_j$ has load $k$.
Consequently, each used strategy $P_i$ has cost $k^{d+1} + \gamma = mk^d$, which is equal to the cost of the alternative strategy $A_i$, so $f$ is a Nash equilibrium.

Next consider a flow $f'$ that assigns the entire demand of every population $i$ to $A_i$.
Then every $a_j$ has load $m$, so each used strategy $A_i$ has cost $m^{d+1}$, while the alternative $P_i$ has cost $km^d + \gamma$.
Using $(m/k)^d = 1 + \omega_d(\eps/(1 + \eps))$, we have
\[
1 - \frac{km^d+\gamma}{m^{d+1}} = (1 - k/m)\bigl(1 - (k/m)^d\bigr) = \lambda_d\bigl((m/k)^d - 1\bigr) = \frac{\eps}{1+\eps}.
\]
Thus $m^{d+1} = (1 + \eps)(km^d + \gamma)$, and $f'$ is an $\eps$-NE.
Since all populations have the same cost within each flow, we have
\[
\frac{C(f')}{C(f)} = \frac{m^{d+1}}{mk^d} = \left(\frac{m}{k}\right)^d = 1 + \omega_d\left(\frac{\eps}{1+\eps}\right),
\]
which exactly matches the upper bound in the first regime of Theorem~\ref{thm:general_polynomial_upperbound}.
\end{example}

The second and third regimes coincide with the approximate PoA bounds of Christodoulou et~al.~\cite{ChristodoulouKS11}.
This is because their tight instances have the same structure as Example~\ref{ex:general_polynomial_upperbound_tight} and only differ in the choice of $k, m, \gamma$.
They use the flow $f$ to upper bound the optimal social cost, and in Appendix~\ref{sec:cks_transfer} we verify that $f$ is actually a Nash equilibrium under their choice of parameters exactly when $\eps \ge \eps_1$.
Thus, while the approximate PoA compares an $\eps$-NE against the socially optimal flow and our $\worst$ compares against the Nash equilibrium, their tight instances remain tight in our setting.

Similar to $\best$, the tight upper bound in the first regime can be achieved for single-commodity nonatomic routing games on Braess graphs for a sequence of values of $\eps$ tending to $0$.
The construction and its analysis can be found in Appendix~\ref{sec:additional_tight_instances}.

\section*{Acknowledgements}
An initial version of this paper was submitted to a conference in February 2026.
That version contained the paper's main idea of comparing the social costs of approximate and exact Nash equilibria in nonatomic routing games.
For degree-$d$ polynomial latencies, it established that the relative change in social cost of $\eps$-approximate equilibria is $O(\sqrt{d\eps})$.
However, the lower-order terms in the bounds were not tight, and the proofs were more complicated.

Beginning in June 2026, the authors used ChatGPT, Claude, and Gemini to help strengthen the lower and upper bounds in Sections~\ref{sec:best} and~\ref{sec:worst} and simplify their proofs, leading to the tight bounds presented in this version.
The authors also used these AI tools to assist with surveying related work and polishing the text.
All generated content has been reviewed, verified, and finalized by the authors, who take full responsibility for the content of the paper.

\bibliographystyle{alpha}
\bibliography{main}

\newpage
\appendix

\section{Properties of the Functions \texorpdfstring{$\mu_d$}{μ\_d} and \texorpdfstring{$\lambda_d$}{λ\_d}}
\label{sec:best_deferred}

In this section we provide the proofs of the properties of the functions $\mu_d$ and $\lambda_d$, which are used in the proof of Theorem~\ref{thm:general_polynomial_meta_lowerbound} and Theorem~\ref{thm:general_polynomial_upperbound}, respectively.

Recall that for integer $d \ge 1$ and real number $x \in [0,1]$, we define $\mu_d(x) = x\left(1 - (1 - x)^{1/d}\right)$.

\lemmuoptima*

\begin{proof}
Continuity on $[0,1]$ is immediate.
For $x \in (0,1)$, direct differentiation gives
\[
\mu_d'(x) = 1 - (1 - x)^{1/d} + (1/d)x(1 - x)^{1/d-1} \ge (1/d)x(1 - x)^{1/d-1} \ge 0,
\]
and the last inequality is strict.
Thus $\mu_d$ is strictly increasing on $(0,1)$ and, by continuity, on $[0,1]$.
Differentiating once more on $(0,1)$ gives
\[
\mu_d''(x) = (1/d)(1 - x)^{1/d-2}(2 - (1 + 1/d)x) \ge 0,
\]
because $2 - (1 + 1/d)x > 0$ for $0 < x < 1$.
Hence $\mu_d$ is convex on $(0,1)$, and continuity extends convexity to $[0,1]$.
Finally, $\mu_d(0) = 0$ and $\mu_d(1) = 1$, so strict monotonicity and continuity imply $\mu_d([0,1]) = [0,1]$.

For the second part, let $F_d(x) = 1 - x + \mu_d(x) = 1 - x(1 - x)^{1/d}$, then
\[
F'_d(x) = -(1 - x)^{1/d-1} \left(1 - \frac{d+1}{d}x\right).
\]
Thus $F'_d(x) < 0$ for $x < d/(d + 1)$, $F'_d(x) > 0$ for $x > d/(d + 1)$, and $F'_d(d/(d + 1)) = 0$.
The claimed monotonicity and uniqueness of the minimizer follow.
\end{proof}

Recall that for integer $d \ge 1$ and real number $x \in [0,+\infty)$, we define
\[
\lambda_d(x) = \frac{x}{1+x}\left(1 - (1 + x)^{-1/d}\right).
\]

\begin{lemma}
\label{lem:lambda_monotonicity}
For every integer $d \ge 1$, $\lambda_d$ is continuous and strictly increasing on $[0,+\infty)$, and maps $[0,+\infty)$ onto $[0,1)$.
\end{lemma}
\begin{proof}
Continuity is immediate.
For $x > 0$,
\[
\lambda_d'(x) = \frac{1-(1+x)^{-1/d}}{(1+x)^2} + \frac{x}{d}(1 + x)^{-1/d-2} \ge 0,
\]
and the derivative is strictly positive.
Therefore $\lambda_d$ is strictly increasing on $[0,+\infty)$.
Finally, $\lambda_d(0) = 0$ and $\lambda_d(x) \to 1$ as $x \to +\infty$, so $\lambda_d([0,+\infty)) = [0,1)$.
\end{proof}

\section{Properties of Series-Parallel Networks}
\label{sec:series_parallel_deferred}

In this section we prove some properties of Nash flows and $\eps$-NE in single-commodity routing on series-parallel networks, which can be used to derive stronger bounds on $\best$ and $\worst$ than the general case.

For a component $H$ and a flow $g$, let $\tau_H(g)$ be the amount of flow routed across $H$.
Fix a Nash flow $f$, and let $L_H$ be the latency of a shortest $s_H$-$t_H$ path at the loads induced by $f$.
If $H$ carries positive flow under $f$, every $s_H$-$t_H$ subpath used by $f$ must have latency $L_H$.
Consequently, if $H$ is a series composition of $H_1, H_2$, then $L_H = L_{H_1} + L_{H_2}$; if $H$ is a parallel composition, then $L_{H_i} \ge L_H$ for each child $H_i$, with equality whenever $\tau_{H_i}(f) > 0$.

\lemseriesparallelusedpathlowerbound*

\begin{proof}
Let $L'$ be the shortest-path latency under $f'$ and set $\delta = \eps L'$.
Every used $s_G$-$t_G$ path has latency at most $L' + \delta$, whereas every $s_G$-$t_G$ path has latency at least $L'$.
It follows that, inside any component carrying positive flow under $f'$, a used subpath has latency at most $\delta$ more than any alternative subpath.

We prove the following claim by induction on the recursive construction of the series-parallel network: for every component $H$ satisfying $\tau_H(f') \ge \tau_H(f)$ and $\tau_H(f') > 0$, the longest used $s_H$-$t_H$ subpath under $f'$ has latency at least $L_H - (h_H - 1)\delta$.

In the base case, if $H$ is a single edge, since we assume that its load under $f'$ is at least its load under $f$, the latency of the edge under $f'$ is at least $L_H$ by monotonicity of latency functions.

If $H$ is a series composition of $H_1, H_2$, then
\[
\tau_{H_1}(f') = \tau_{H_2}(f') = \tau_H(f') \ge \tau_H(f) = \tau_{H_1}(f) = \tau_{H_2}(f).
\]
By inductive hypothesis on $H_1, H_2$, the longest used path in $H_1$ has latency at least $L_{H_1} - (h_{H_1} - 1)\delta$, and the longest used path in $H_2$ has latency at least $L_{H_2} - (h_{H_2} - 1)\delta$.
Thus the shortest path in $H_2$ has latency at least $L_{H_2} - h_{H_2}\delta$.
Therefore, for a used path in $H$ whose subpath in $H_1$ is a longest used path in $H_1$, its latency is at least
\[
L_{H_1} - (h_{H_1} - 1)\delta + L_{H_2} - h_{H_2}\delta = L_H - (h_H - 1)\delta,
\]
which follows from $L_H = L_{H_1} + L_{H_2}$ and $h_H = h_{H_1} + h_{H_2}$.

Otherwise, if $H$ is a parallel composition of $H_1, H_2$, then since $\tau_H(f') \ge \tau_H(f)$, there exists $i \in \{1,2\}$ such that $\tau_{H_i}(f') \ge \tau_{H_i}(f)$ and $\tau_{H_i}(f') > 0$.
The inductive hypothesis gives a used path in $H_i$ of latency at least
\[
L_{H_i} - (h_{H_i} - 1)\delta \ge L_H - (h_H - 1)\delta,
\]
since $L_{H_i} \ge L_H$ and $h_{H_i} \le h_H$.
This concludes the proof of the claim.

Finally, apply the claim to the full network $G$: the longest used path under $f'$ has latency at least $L_G - (h_G - 1)\delta$.
But we also know that its latency is at most $L' + \delta$.
Therefore,
\[
L' \ge L_G - (h_G - 1)\delta - \delta = L_G - h_G\delta = L_G - h_G\eps L',
\]
which implies $L' \ge L_G/(1 + h_G\eps)$.
\end{proof}

\begin{lemma}
\label{lem:series_parallel_path_upperbound}
Let $G$ be a series-parallel network with monotone latency functions.
Let $f$ be a Nash flow, and let $L_G$ be the common latency of the paths used by $f$.
Then for every flow $f'$, some $s_G$-$t_G$ path has latency at most $L_G$ under $f'$.
\end{lemma}
\begin{proof}
We prove the following claim by induction: if a component $H$ satisfies $0 < \tau_H(f)$ and $\tau_H(f') \le \tau_H(f)$, then it contains an $s_H$-$t_H$ path whose latency under $f'$ is at most $L_H$.
The lemma follows from applying the claim to the full network.

The base case where $H$ is a single edge follows from monotonicity of the latency function.

If $H$ is a series composition of $H_1, H_2$, then
\[
\tau_{H_1}(f') = \tau_{H_2}(f') = \tau_H(f') \le \tau_H(f) = \tau_{H_1}(f) = \tau_{H_2}(f),
\]
so by inductive hypothesis, there exists an $s_{H_i}$-$t_{H_i}$ path with latency at most $L_{H_i}$ for $i = 1, 2$.
Concatenating these two paths gives an $s_H$-$t_H$ path of latency at most $L_{H_1} + L_{H_2} = L_H$.

If $H$ is a parallel composition of $H_1, H_2$, then there exists $i \in \{1,2\}$ with $\tau_{H_i}(f') \le \tau_{H_i}(f)$ and $\tau_{H_i}(f) > 0$, so by inductive hypothesis, there exists an $s_{H_i}$-$t_{H_i}$ path (which is also an $s_H$-$t_H$ path) with latency at most $L_{H_i} = L_H$.
\end{proof}

In single-commodity routing games on series-parallel networks, the upper bound on $\worst$ is $1 + \eps$.
This result was first shown by Kleer and Sch\"{a}fer~\cite{sagt/KleerS17}.
We give a self-contained proof for completeness.

\begin{theorem}
\label{thm:series_parallel_upperbound}
Consider any nonatomic routing game on a series-parallel network with monotone latency functions.
Let $f$ be a Nash flow.
Then for every $\eps \ge 0$, every $\eps$-NE $f'$ satisfies $\frac{C(f')}{C(f)} \le 1 + \eps$.
\end{theorem}
\begin{proof}
Let $r$ be the demand and let $L_G$ be the common latency of the paths used by $f$, so $C(f) = rL_G$.
By Lemma~\ref{lem:series_parallel_path_upperbound}, some path has latency at most $L_G$ under $f'$.
Since $f'$ is an $\eps$-NE, every path used by $f'$ has latency at most $(1 + \eps)L_G$, so $C(f') \le (1 + \eps)rL_G = (1 + \eps)C(f)$.
\end{proof}

The bound is tight even for a Pigouvian network with constant latency functions.

\begin{example}
\label{ex:sp-upper-tight}
Consider a Pigouvian network where the upper edge has constant latency $1 + \eps$ and the lower edge has constant latency $1$.
The Nash flow routes one unit of traffic on the lower edge and has social cost $1$.
The flow $f'$ that routes all traffic on the upper edge is an $\eps$-NE and has social cost $1 + \eps$.
\end{example}

\section{Additional Tight Instances}
\label{sec:additional_tight_instances}

\subsection{Tight Instances using Braess Graphs}
\label{sec:braess_tight_examples}

Braess graphs have long served as canonical examples of extreme behavior in selfish routing.
The next two examples show that the same topology can essentially achieve the tight bound for both $\best$ and $\worst$.

\begin{example}[Best $\eps$-NE]
\label{ex:braess_polynomial_lowerbound_tight}
Fix integers $d$ and $k$ such that $\theta_k := 1 - k^d/(k + 1)^d \le d/(d + 1)$.
Construct a network $B^d_k = (V_k,E_k)$ with vertex set $V_k = \{s,v_1,\ldots,v_k,w_1,\ldots,w_k,t\}$ and the following edges:
\begin{itemize}
    \item $(s,w_1)$ and $(v_k,t)$, with latency $\ell(x) = 1/(k + 1)$;
    \item $(s,v_i)$, with latency $\ell(x) = i k^d x^d/(k + 1)$, for $1 \le i \le k$;
    \item $(v_i,w_i)$, with latency $\ell(x) = 0$, for $1 \le i \le k$;
    \item $(v_i,w_{i+1})$, with latency $\ell(x) = 1/(k + 1)$, for $1 \le i \le k - 1$;
    \item $(w_i,t)$, with latency $\ell(x) = (k - i + 1)k^d x^d/(k + 1)$, for $1 \le i \le k$.
\end{itemize}

As an example, Figure~\ref{fig:braess_polynomial} illustrates the network and latency functions of $B_3^d$.

\begin{figure}[ht]
\centering
\begin{tikzpicture}[scale=0.8, auto, thick, >=Stealth]
    \node[node] (s) at (-7, 0) {$s$};
    \node[node] (v1) at (-2, 1) {$v_1$};
    \node[node] (v2) at (-2, -1) {$v_2$};
    \node[node] (v3) at (0, -3) {$v_3$};
    \node[node] (w1) at (0, 3) {$w_1$};
    \node[node] (w2) at (2, 1) {$w_2$};
    \node[node] (w3) at (2, -1) {$w_3$};
    \node[node] (t) at (7, 0) {$t$};

    \draw[->] (s) edge[bend left=20] node[midway, above, black] {$\frac{1}{4}$} (w1);
    \draw[->] (s) edge node[pos=0.6, above, black] {$\frac{3^d}{4}x^d$} (v1);
    \draw[->] (s) edge node[pos=0.6, above, black] {$\frac{2\cdot3^d}{4}x^d$} (v2);
    \draw[->] (s) edge[bend right=20] node[pos=0.6, above, black] {$\frac{3\cdot3^d}{4}x^d$} (v3);
    \draw[->] (v1) edge node[midway, above, black] {$0$} (w1);
    \draw[->] (v2) edge node[midway, above, black] {$0$} (w2);
    \draw[->] (v3) edge node[midway, above, black] {$0$} (w3);
    \draw[->] (v1) edge node[midway, above, black] {$\frac{1}{4}$} (w2);
    \draw[->] (v2) edge node[midway, above, black] {$\frac{1}{4}$} (w3);
    \draw[->] (w1) edge[bend left=20] node[pos=0.4, above, black] {$\frac{3\cdot3^d}{4}x^d$} (t);
    \draw[->] (w2) edge node[pos=0.4, above, black] {$\frac{2\cdot3^d}{4}x^d$} (t);
    \draw[->] (w3) edge node[pos=0.4, above, black] {$\frac{3^d}{4}x^d$} (t);
    \draw[->] (v3) edge[bend right=20] node[midway, above, black] {$\frac{1}{4}$} (t);
\end{tikzpicture}
\caption{Illustration of $B_3^d$.}
\label{fig:braess_polynomial}
\end{figure}
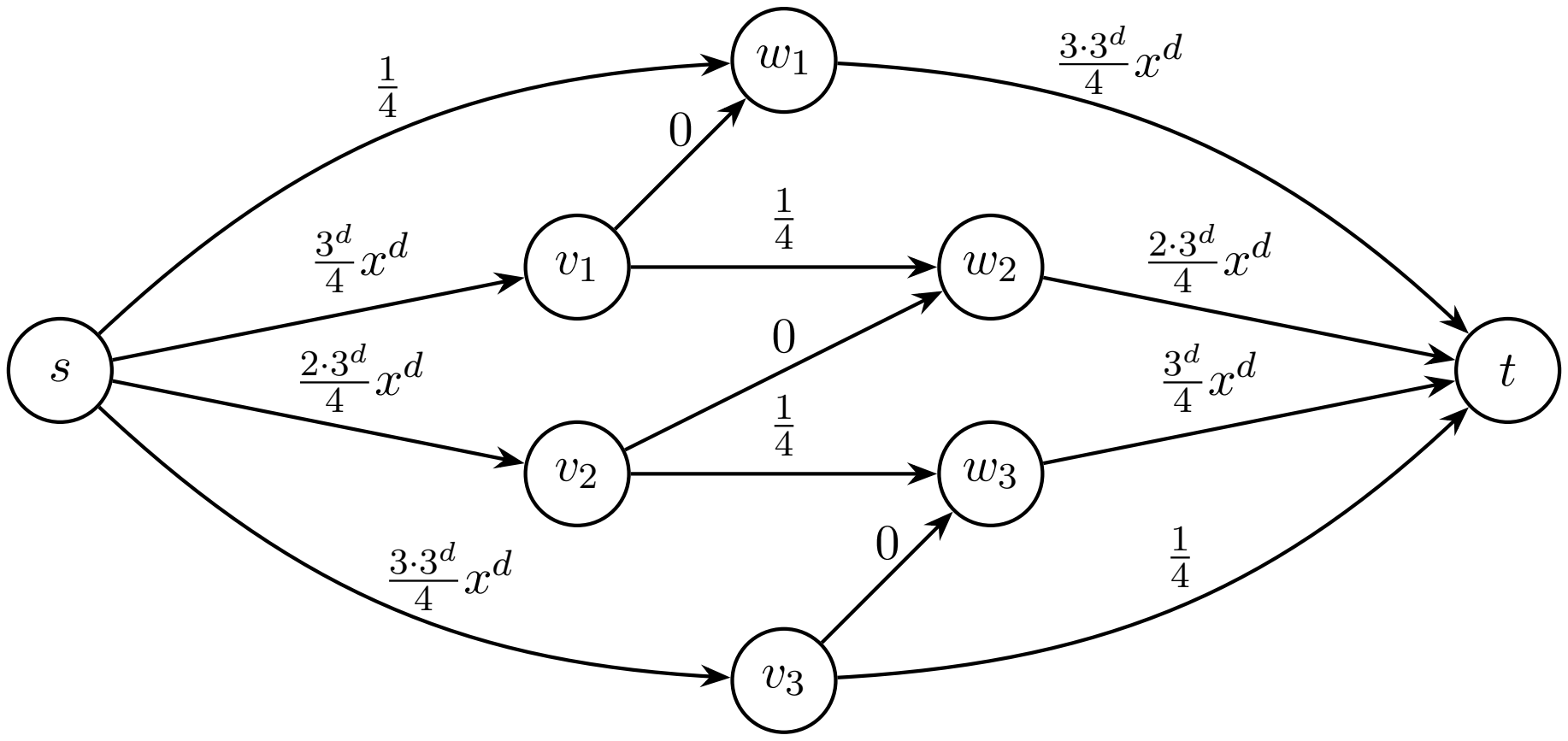

The Nash flow $f$ routes $1/k$ units of flow on each path $s \to v_i \to w_i \to t$, for $1 \le i \le k$.
Every such path has latency
\[
\frac{ik^d}{k+1}\cdot\frac{1}{k^d} + 0 + \frac{(k-i+1)k^d}{k+1}\cdot\frac{1}{k^d} = 1.
\]
Every unused path also has latency $1$, so $f$ is a Nash flow and $C(f) = 1$.

Now consider a flow $f'$ that routes $1/(k + 1)$ units on each of the following $k + 1$ paths:
\begin{itemize}
    \item $s \to w_1 \to t$;
    \item $s \to v_k \to t$;
    \item $s \to v_i \to w_{i+1} \to t$, for each $1 \le i \le k - 1$.
\end{itemize}
Every resource with nonzero latency on one of these paths has load $1/(k + 1)$.
Hence each used path has latency
\[
\frac{1}{k+1} + \frac{k\cdot k^d}{k+1}\cdot\frac{1}{(k+1)^d} = \frac{(k+1)^d+k^{d+1}}{(k+1)^{d+1}}.
\]
The shortest paths under $f'$ are $s \to v_i \to w_i \to t$, which have latency
\[
\frac{ik^d}{k+1}\cdot\frac{1}{(k+1)^d} + 0 + \frac{(k-i+1)k^d}{k+1}\cdot\frac{1}{(k+1)^d} = \frac{(k+1)k^d}{(k+1)^{d+1}}.
\]
The ratio between the used-path latency and the shortest-path latency is
\[
\frac{(k+1)^d+k^{d+1}}{(k+1)k^d} = 1 + \frac{(k+1)^d-k^d}{(k+1)k^d}.
\]
Therefore, $f'$ is an $\eps_k$-NE, where
\[
\eps_k = \frac{(k+1)^d-k^d}{(k+1)k^d},\qquad
C(f') = \frac{(k+1)^d+k^{d+1}}{(k+1)^{d+1}}.
\]
By the definition of $\theta_k$,
\[
\mu_d(\theta_k) = \frac{(k+1)^d-k^d}{(k+1)^{d+1}} = \eps_k(1 - \theta_k),
\]
so $\rho_d(\eps_k) = \theta_k$.
Our choice of $k$ ensures that $\theta_k$ is the active term in Corollary~\ref{cor:general_polynomial_lowerbound}.
Finally,
\[
\frac{C(f')}{C(f)} = \frac{(k+1)^d+k^{d+1}}{(k+1)^{d+1}} = \left(\frac{k}{k+1}\right)^d + \frac{(k+1)^d-k^d}{(k+1)^{d+1}} = 1 - \theta_k + \mu_d(\theta_k).
\]
Thus the bound in Corollary~\ref{cor:general_polynomial_lowerbound} is attained exactly along a sequence $\eps_k \to 0$.

The same construction also establishes the claim for arbitrary monotone latencies in Section~\ref{sec:monotone_lower}.
Let $d = \lfloor k\ln k/2\rfloor$.
This choice satisfies $\theta_k \le d/(d + 1)$ for all sufficiently large $k$.
As $k \to \infty$, we have $(1 + 1/k)^d \sim \sqrt{k}$, so
\[
\eps_k = \frac{(1 + 1/k)^d - 1}{k + 1} \to 0,
\qquad
\frac{C(f')}{C(f)} = \frac{1}{k + 1} + \left(\frac{k}{k + 1}\right)^{d+1} \to 0.
\]
For any fixed $\eps > 0$, we have $\eps_k \le \eps$ for all sufficiently large $k$, so the corresponding flows are $\eps$-NE with cost ratios tending to $0$.
\end{example}

\begin{example}[Worst $\eps$-NE]
\label{ex:braess_polynomial_upperbound_tight}
Fix an integer $d \ge 1$.
For every sufficiently large integer $k$, construct $\widehat B^d_k$ on the same vertex and edge sets as $B_k^d$, with the following latency functions:
\begin{itemize}
    \item $(s,w_1)$ and $(v_k,t)$ have latency $\ell(x) = 1/(k + 1)$;
    \item $(s,v_i)$ has latency $\ell(x) = i(k + 1)^{d-1}x^d$, for $1 \le i \le k$;
    \item $(v_i,w_i)$ has latency $\ell(x) = 0$, for $1 \le i \le k$;
    \item $(v_i,w_{i+1})$ has latency $\ell(x) = 1/(k + 1)$, for $1 \le i \le k - 1$;
    \item $(w_i,t)$ has latency $\ell(x) = (k - i + 1)(k + 1)^{d-1}x^d$, for $1 \le i \le k$.
\end{itemize}

Figure~\ref{fig:braess_polynomial_upper} illustrates the network and latency functions of $\widehat B_3^d$.

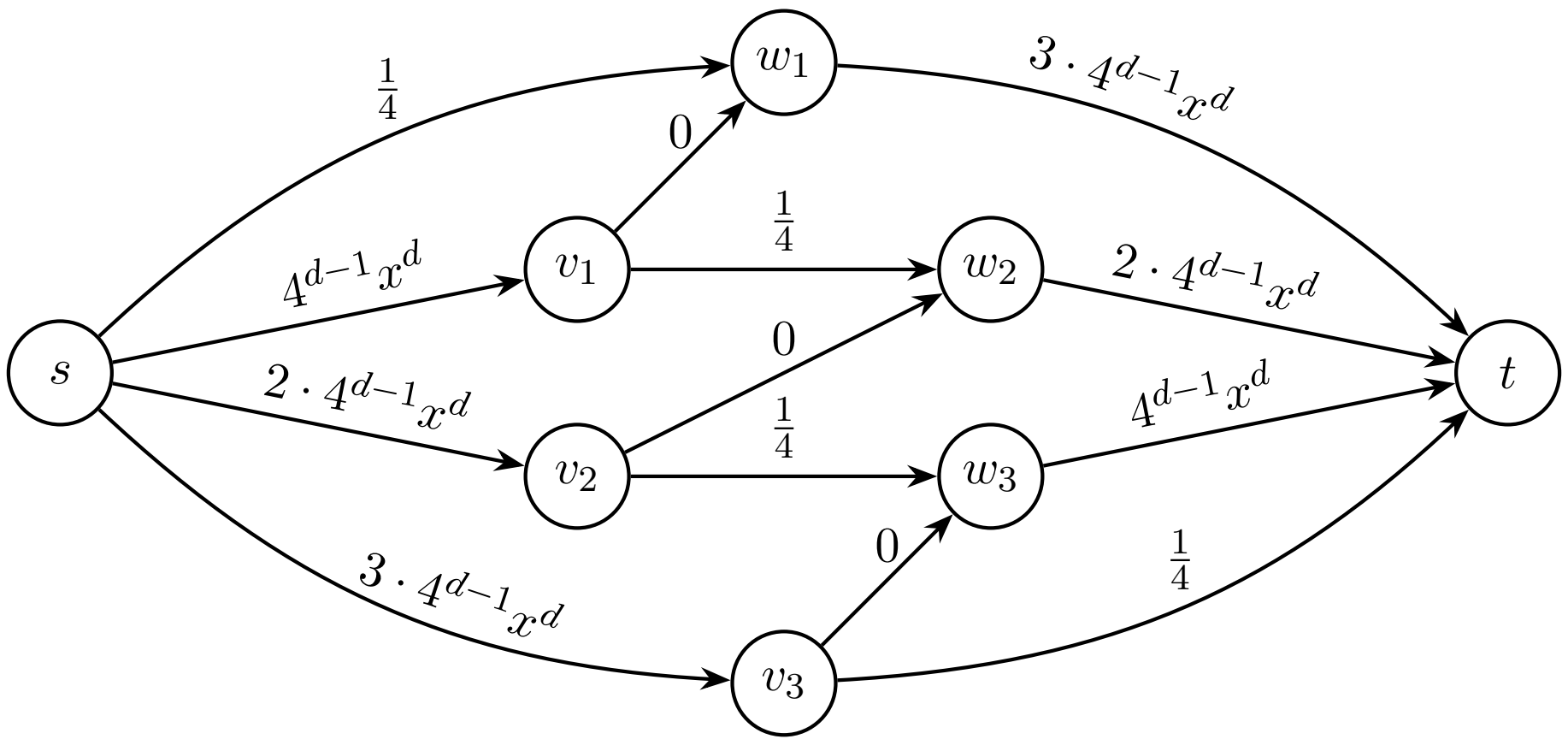
\begin{figure}[ht]
\centering
\begin{tikzpicture}[scale=0.8, auto, thick, >=Stealth]
    \node[node] (s) at (-7, 0) {$s$};
    \node[node] (v1) at (-2, 1) {$v_1$};
    \node[node] (v2) at (-2, -1) {$v_2$};
    \node[node] (v3) at (0, -3) {$v_3$};
    \node[node] (w1) at (0, 3) {$w_1$};
    \node[node] (w2) at (2, 1) {$w_2$};
    \node[node] (w3) at (2, -1) {$w_3$};
    \node[node] (t) at (7, 0) {$t$};

    \draw[->] (s) edge[bend left=20] node[midway, above, black] {$\frac{1}{4}$} (w1);
    \draw[->] (s) edge node[pos=0.6, sloped, above, black] {$4^{d-1}x^d$} (v1);
    \draw[->] (s) edge node[pos=0.6, sloped, above, black] {$2\cdot4^{d-1}x^d$} (v2);
    \draw[->] (s) edge[bend right=20] node[pos=0.6, sloped, above, black] {$3\cdot4^{d-1}x^d$} (v3);
    \draw[->] (v1) edge node[midway, above, black] {$0$} (w1);
    \draw[->] (v2) edge node[midway, above, black] {$0$} (w2);
    \draw[->] (v3) edge node[midway, above, black] {$0$} (w3);
    \draw[->] (v1) edge node[midway, above, black] {$\frac{1}{4}$} (w2);
    \draw[->] (v2) edge node[midway, above, black] {$\frac{1}{4}$} (w3);
    \draw[->] (w1) edge[bend left=20] node[pos=0.4, sloped, above, black] {$3\cdot4^{d-1}x^d$} (t);
    \draw[->] (w2) edge node[pos=0.4, sloped, above, black] {$2\cdot4^{d-1}x^d$} (t);
    \draw[->] (w3) edge node[pos=0.4, sloped, above, black] {$4^{d-1}x^d$} (t);
    \draw[->] (v3) edge[bend right=20] node[midway, above, black] {$\frac{1}{4}$} (t);
\end{tikzpicture}
\caption{Illustration of $\widehat B_3^d$.}
\label{fig:braess_polynomial_upper}
\end{figure}

The Nash flow $f$ routes $1/(k + 1)$ units on each of the following $k + 1$ paths:
\begin{itemize}
    \item $s \to w_1 \to t$;
    \item $s \to v_k \to t$;
    \item $s \to v_i \to w_{i+1} \to t$, for each $1 \le i \le k - 1$.
\end{itemize}
Every used path has latency
\[
\frac{1}{k+1} + k(k + 1)^{d-1}\cdot\frac{1}{(k+1)^d} = 1.
\]
Moreover, the paths $s \to v_i \to w_i \to t$, for $1 \le i \le k$, have latency
\[
i(k + 1)^{d-1}\cdot\frac{1}{(k+1)^d} + 0
+ (k - i + 1)(k + 1)^{d-1}\cdot\frac{1}{(k+1)^d} = 1.
\]
Thus $f$ is a Nash flow and $C(f) = 1$.

Now consider a flow $f'$ that routes $1/k$ units on each path $s \to v_i \to w_i \to t$, for $1 \le i \le k$.
Each used path has latency
\[
i(k + 1)^{d-1}\cdot\frac{1}{k^d} + 0
+ (k - i + 1)(k + 1)^{d-1}\cdot\frac{1}{k^d}
= \frac{(k+1)^d}{k^d}.
\]
Under $f'$, the shortest paths are $s \to w_1 \to t$, $s \to v_k \to t$, and $s \to v_i \to w_{i+1} \to t$ for $1 \le i \le k - 1$.
Each has latency
\[
\frac{1}{k+1} + k(k + 1)^{d-1}\cdot\frac{1}{k^d}
= \frac{1}{k+1} + \frac{(k+1)^{d-1}}{k^{d-1}}.
\]
Therefore, $f'$ is an $\eps_k$-NE with
\[
1 + \eps_k = \frac{(k+1)^d/k^d}{\frac{1}{k+1}+\frac{(k+1)^{d-1}}{k^{d-1}}},
\qquad
\frac{C(f')}{C(f)} = \frac{(k+1)^d}{k^d}.
\]
Writing $\alpha = (k + 1)^d/k^d$, we obtain
\[
\frac{\eps_k}{1+\eps_k}
= 1 - \frac{1}{1+\eps_k}
= 1 - \frac{\frac{1}{k+1}+\frac{(k+1)^{d-1}}{k^{d-1}}}{\frac{(k+1)^d}{k^d}}
= \left(1 - \alpha^{-1}\right)\left(1 - \alpha^{-1/d}\right)
= \lambda_d(\alpha - 1).
\]
Thus $\alpha = 1 + \omega_d(\eps_k/(1 + \eps_k))$.
So the first regime of Theorem~\ref{thm:general_polynomial_upperbound} is attained along a sequence $\eps_k \to 0$.
\end{example}

\subsection{Tight Instances for the Second and Third Regimes of Theorem~\ref{thm:general_polynomial_upperbound}}
\label{sec:cks_transfer}

We use exactly the same instances as in Christodoulou et~al.~\cite{ChristodoulouKS11}.
The instances have the same combinatorial structure as Example~\ref{ex:general_polynomial_upperbound_tight}, and we use the same choices of $f$ and $f'$.
The only difference is in the choice of the constant latency $\gamma$ for the private resources.

\begin{example}
Use the two flows $f, f'$ from Example~\ref{ex:general_polynomial_upperbound_tight}.
For each population, the used and alternative strategies under $f$ have costs $k^{d+1} + \gamma$ and $mk^d$, respectively, while those under $f'$ have costs $m^{d+1}$ and $km^d + \gamma$.
We specify $m \ge k \ge 1$ and $\gamma \ge 0$ separately for the two regimes.

\paragraph{Second regime: $\eps_1 \le \eps \le \eps_2$.}
Take integers $m \ge k \ge 1$ such that $m/k$ gives an arbitrarily good approximation to $(d + 1)^{1/d}$.
Below we simply assume that $m/k = (d + 1)^{1/d}$.
To make the ratio between the used and alternative costs under $f'$ equal to $1 + \eps$, set $\gamma = m^{d+1}/(1 + \eps) - km^d$.
We now determine when this choice gives a valid instance with $f$ a Nash equilibrium.

First, the requirement $\gamma \ge 0$ is equivalent to $1 + \eps \le m/k$, or $\eps \le (d + 1)^{1/d} - 1 = \eps_2$.
Second, $f$ is a Nash equilibrium if and only if $k^{d+1} + \gamma \le mk^d$.
Substituting our choice of $\gamma$ and rearranging gives
\[
\eps \ge \frac{m^{d+1}}{km^d+(m-k)k^d} - 1 = \frac{d\bigl((d+1)^{1/d}-1\bigr)}{d+(d+1)^{1/d}} = \eps_1.
\]

Thus the two requirements hold exactly when $\eps_1 \le \eps \le \eps_2$.
In this range, $f$ is a Nash equilibrium and $f'$ is an $\eps$-NE.
Since all populations have the same cost within each flow, we have
\[
\frac{C(f')}{C(f)} = \frac{m^{d+1}}{k^{d+1}+\gamma} = \left(\frac{1}{1+\eps} - \frac{k}{m} + \left(\frac{k}{m}\right)^{d+1}\right)^{-1} = \left(\frac{1}{1+\eps} - c_d\right)^{-1},
\]
which exactly matches the upper bound in the second regime of Theorem~\ref{thm:general_polynomial_upperbound}.

\paragraph{Third regime: $\eps \ge \eps_2$.}
Take integers $m \ge k \ge 1$ such that $m/k$ gives an arbitrarily good approximation to $1 + \eps$.
Below we simply assume that $m/k = 1 + \eps$.
Simply set $\gamma = 0$.

Since $k^{d+1} \le mk^d$, the flow $f$ is a Nash equilibrium.
Moreover, $m^{d+1}/(km^d) = m/k = 1 + \eps$, so $f'$ is an $\eps$-NE.
Finally,
\[
\frac{C(f')}{C(f)} = \frac{m^{d+1}}{k^{d+1}} = \left(\frac{m}{k}\right)^{d+1} = (1 + \eps)^{d+1},
\]
which exactly matches the upper bound in the third regime of Theorem~\ref{thm:general_polynomial_upperbound}.
\end{example}

\end{document}